\documentclass[a4paper,fleqn,usenatbib]{mnras}

\usepackage{newtxtext,newtxmath}
\usepackage[T1]{fontenc}
\usepackage{ae,aecompl}

\usepackage{graphicx}	% Including figure files
\usepackage{amsmath}	% Advanced maths commands  %To put text mode inside math mode
\usepackage{newtxtext,newtxmath}
\usepackage[flushleft]{threeparttable}
\usepackage[pdftex,usenames,dvipsnames]{xcolor}
\usepackage{natbib}
\usepackage{aas_macros}
\usepackage{url}
\usepackage[none]{hyphenat}
\usepackage{times}
\usepackage{array}
\title[MAD]{MaNGA AGN dwarf galaxies (MAD) - I. A new sample of AGN in dwarf galaxies with spatially resolved spectroscopy}

\author[Mezcua et al.]{
M. Mezcua$^{1,2}$\thanks{E-mail: marmezcua.astro@gmail.com}, H. Dom\'inguez S\'anchez$^{3}$
\\
$^{1}$ Institute of Space Sciences (ICE, CSIC), Campus UAB, Carrer de Magrans, 08193 Barcelona, Spain\\
$^{2}$ Institut d'Estudis Espacials de Catalunya (IEEC), Carrer Gran Capit\`{a}, 08034 Barcelona, Spain\\
${3}$  Centro de Estudios de Física del Cosmos de Aragón (CEFCA), Plaza San Juan, 1, 44001, Teruel, Spain\\
}

\date{Accepted 2024 January 24. Received 2024 January 8; in original form 2023 May 19}

\pubyear{XXX}

\begin{document}
\label{firstpage}
\pagerange{\pageref{firstpage}--\pageref{lastpage}}
\maketitle

\begin{abstract}
The finding of active galactic nuclei (AGN) in dwarf galaxies has important implications for galaxy evolution and supermassive black hole formation models. Yet, how AGN in dwarf galaxies form is still debated, in part due to scant demographics. We make use of the MaNGA survey, comprising $\sim$10,000 galaxies at z $<$ 0.15, to identify AGN dwarf galaxies using a spaxel by spaxel classification in three spatially-resolved emission line diagnostic diagrams (the [NII-, [SII]- and [OI]-BPT) and the WHAN diagram. This yields a sample of 664 AGN dwarf galaxies, the largest to date, and an AGN fraction of $\sim20\%$ that is significantly larger than that of single-fiber-spectroscopy studies (i.e. $\sim1\%$). This can be explained by the lower bolometric luminosity ($< 10^{42}$ erg s$^{-1}$) and accretion rate (sub-Eddington) of the MaNGA AGN dwarf galaxies. We additionally identify 1,176 SF-AGN (classified as star-forming in the [NII]-BPT but as AGN in the [SII]- and [OI]-BPT), 122 Composite, and 173 LINER sources. The offset between the optical center of the galaxy and the median position of the AGN spaxels is more than 3 arcsec for $\sim$62\% of the AGN, suggesting that some could be off-nuclear. We also identify seven new broad-line AGN with log $M_\mathrm{BH}$ = 5.0 - 5.9 $M_\mathrm{\odot}$. Our results show how integral-field spectroscopy is a powerful tool for uncovering faint and low-accretion AGN and better constraining the demographics of AGN in dwarf galaxies.
\end{abstract}


%%%%%%%%%%%%%%%%%%%%%%%%%%%%%%%%%%%%%%%%%%%%%%%%%%

%%%%%%%%%%%%%%%%% BODY OF PAPER %%%%%%%%%%%%%%%%%%

\section{Introduction}

The investigation of active galactic nuclei (AGN) in dwarf galaxies (stellar mass $M_\mathrm{*} \leq 3 \times 10^{9}$ M$_{\odot}$) has become of major importance for studies of black hole (BH) formation and galaxy evolution. On one hand, dwarf galaxies have low mass, low metallicity, and a presumably modest merger history, hence they are thought to resemble the first galaxies formed in the early Universe. On the other hand, these primordial dwarf galaxies are candidate to host BHs of mass M$_\mathrm{BH} \sim 100-10^{6}$ M$_{\odot}$ (also called intermediate-mass BHs, IMBHs; e.g., \citealt{2017IJMPD..2630021M}; \citealt{2020ARA&A..58..257G}) formed even earlier on (at redshifts z $>$ 10) and that are the seeds of the supermassive BHs (SMBHs; $M_\mathrm{BH} > 10^{6}$ M$_{\odot}$) detected at z$\sim$6-7 (e.g., \citealt{2021ApJ...907L...1W}) and even at z$\sim$10 thanks to the superb sensitivity of the \textit{James Webb Space
Telescope} (e.g., \citealt{2023arXiv230512492M}). Those seed BHs that did not become supermassive should be found in today's dwarf galaxies as relic IMBHs if dwarf galaxies are regulated by supernova feedback, which expels the gas from the center via stellar winds and prevents BH growth (e.g., \citealt{2010MNRAS.408.1139V}; \citealt{2017MNRAS.468.3935H}). In this scenario, deriving the AGN occupation fraction (as proxy of BH occupation fraction) and luminosity function in dwarf galaxies can provide important constraints on the formation mechanisms of the high-z seed BHs (e.g., \citealt{2008MNRAS.383.1079V};  \citealt{2018MNRAS.481.3278R}; \citealt{2022MNRAS.516.2736B}; \citealt{2023MNRAS.523.5610B}). 

A BH occupation fraction of 100\% is expected for dwarf galaxies of log $M_\mathrm{*} \sim 10^{8}-10^{9}$ M$_{\odot}$ if seed BHs of up to 1,000 M$_{\odot}$ formed from the death of the first generation of Population III stars (`light' seeds; e.g., \citealt{2001ApJ...551L..27M}; \citealt{2004ARA&A..42...79B}; see review by e.g., \citealt{2019PASA...36...27W}). A lower BH occupation fraction of $\sim$50\% is instead expected in the case of `heavy' seeds of up to $\sim10^{5}-10^{6}$ M$_{\odot}$ formed from direct collapse of pristine gas (e.g., \citealt{1994ApJ...432...52L}; \citealt{2003ApJ...596...34B}; \citealt{2022Natur.607...48L}). The predicted BH occupation fraction is however strongly dependent on the modelling of cosmological simulations (e.g., \citealt{2022MNRAS.514.4912H}), and distinguishing between light and heavy seeds becomes blurry if heavy BHs are formed at later epochs in nuclear stellar clusters (e.g., \citealt{2021MNRAS.501.1413N}; \citealt{2022ApJ...927..231F}; \citealt{2022ApJ...929...84B}), seed BH growth is significantly impacted by galaxy mergers and AGN feedback (e.g., \citealt{2014ApJ...794..115D}; \citealt{2019NatAs...3....6M,2021IAUS..359..238M}), BHs are off-nuclear (e.g., \citealt{2019MNRAS.482.2913B}; \citealt{2021MNRAS.503.6098R}; \citealt{2023MNRAS.518.1880B}), or other possible seed BH formation mechanisms are considered (e.g., \citealt{2014MNRAS.437.1576B}; \citealt{2020ARNPS..70..355C}; \citealt{2020ApJ...892L...4L}; \citealt{2022MNRAS.512.6192S}).

Multiple studies have yet focused on identifying AGN in dwarf galaxies (see reviews by \citealt{2017IJMPD..2630021M}; \citealt{2020ARA&A..58..257G}; \citealt{2022NatAs...6...26R}), deriving their AGN fraction (e.g., \citealt{2016ApJ...831..203P}; \citealt{2018MNRAS.478.2576M}; \citealt{2020MNRAS.492.2268B}; \citealt{2024MNRAS.527.1962B}), and probing the impact of AGN feedback both from an observational (e.g., \citealt{2018ApJ...861...50B}; \citealt{2019ApJ...884...54M}; \citealt{2019MNRAS.488..685M}; \citealt{2020ApJ...905..166L}; \citealt{2021ApJ...911...70B}; \citealt{2022MNRAS.511.4109D}; \citealt{2022Natur.601..329S}) and theoretical perspective (e.g., \citealt{2018MNRAS.473.5698D}; \citealt{2019MNRAS.487.5549B}; \citealt{2022ApJ...936...82S}; \citealt{2021MNRAS.503.3568K,2022MNRAS.516.2112K}). 
Most searches are based on the use of optical emission line diagnostic diagrams (e.g., [OIII]$\lambda$5007/H$\beta$ versus [NII]$\lambda$6583/H$\alpha$, from now on [NII]-BPT\footnote{\cite{1981PASP...93....5B} - BPT- diagrams}, and [OIII]$\lambda$5007/H$\beta$ versus [SII]$\lambda$6717,6731/H$\alpha$, from now on [SII]-BPT) to distinguish between AGN and star-formation ionisation, and the detection of broad Balmer lines to measure a BH mass $M_\mathrm{BH} \lesssim 10^{6}$ M$_{\odot}$ consistent with that of IMBHs (e.g., \citealt{1989ApJ...342L..11F}; \citealt{2004ApJ...607...90B}; \citealt{2004ApJ...610..722G,2007ApJ...670...92G}; \citealt{2013ApJ...775..116R}; \citealt{2014AJ....148..136M}; \citealt{2015ApJ...809L..14B}; \citealt{2017A&A...602A..28M}; \citealt{2018ApJ...863....1C}). While having provided hundreds of low-mass AGN, these methods are biased towards i) nearby (z$<$0.3; but see \citealt{2019ApJ...885L...3H}; \citealt{2023arXiv230311946H}; \citealt{2023ApJ...954L...4K}; \citealt{2023ApJ...943L...5M}; \citealt{2023MNRAS.518..724S}; \citealt{2023A&A...677A.145U} for higher z) dwarf galaxies with solar to super-solar metallicities (as low-metallicity galaxies have low [NII]/H$\alpha$ values and thus move to the star-forming region of the [NII]-BPT; e.g., \citealt{2006MNRAS.371.1559G}; \citealt{2019ApJ...870L...2C}), and  ii) central AGN with high luminosities (e.g., \citealt{2020ARA&A..58..257G}). These biases can be circumvented by the addition of the [OIII]$\lambda$5007/H$\beta$ versus [OI]$\lambda$6300 diagram (from now on [OI]-BPT), which is insensitive to metallicity (e.g., \citealt{2022ApJ...931...44P}), and the use of X-ray observations (e.g., \citealt{2013ApJ...773..150S}; \citealt{2015ApJ...805...12L}; \citealt{2016ApJ...831..203P}; \citealt{2016ApJ...817...20M,2018MNRAS.478.2576M}; \citealt{2017ApJ...837...48C}; \citealt{2020MNRAS.492.2268B,2022MNRAS.510.4556B}), radio observations (e.g., \citealt{2019MNRAS.488..685M}; \citealt{2020ApJ...888...36R}; \citealt{2022MNRAS.511.4109D}), optical and infrared variability (e.g., \citealt{2018ApJ...868..152B,2020ApJ...896...10B}; \citealt{2020ApJ...889..113M}; \citealt{2020ApJ...900...56S}; \citealt{2022ApJ...936..104W}), or coronal lines (e.g., \citealt{2007ApJ...663L...9S}; \citealt{2018ApJ...861..142C,2020ApJ...895..147C,2021ApJ...912L...2C}; \citealt{2021ApJ...922..155M}; \citealt{2022ApJ...937....7S}; \citealt{2023ApJ...946L..38R}) to identify AGN. X-ray observations offer, in addition, the possibility to derive an AGN fraction corrected for completeness and out to intermediate redshifts (e.g., \citealt{2018MNRAS.478.2576M}; \citealt{2020MNRAS.492.2268B,2022MNRAS.510.4556B}; \citealt{2024MNRAS.527.1962B}), which can then be contrasted to predictions of cosmological simulations (e.g., \citealt{2022MNRAS.514.4912H}).

Both X-ray and radio observations are moreover excellent tools for identifying AGN that are wandering the dwarf galaxy body rather than being central (e.g., \citealt{2018MNRAS.478.2576M}; \citealt{2020ApJ...888...36R}), as predicted by seed BH formation models (e.g., \citealt{2019MNRAS.482.2913B}; \citealt{2019MNRAS.486..101P}). However, both X-ray and radio campaigns are observationally expensive and thus limited in sky coverage. Moreover, only $\sim$10\% of AGN are radio-emitters. In the optical, such wandering or off-nuclear AGN can be identified with the use of integral field unit (IFU) spectroscopy (e.g., \citealt{2022ApJ...927...23C}). \citeauthor{2020ApJ...898L..30M} (2020, MDS+2020 from now on) used the SDSS\footnote{Sloan Digital Sky Survey, \url{https://www.sdss.org}}/MaNGA (Mapping NearbyGalaxies at APO; \citealt{2015ApJ...798....7B}) survey Data Release\footnote{\url{http://www.sdss.org/surveys/manga}} 15 to search for AGN in dwarf galaxies. Out of 1609 dwarves, they found 37 AGN based on spatially-resolved emission line diagnostic diagrams. Most of these 37 AGN dwarf galaxies were offset from the optical center of the galaxy - probably due to the AGN being off-nuclear or switched off or the central emission being dominated by star formation - and hence had been missed by previous searches using single-fiber SDSS spectroscopy. 

We note that all previous searches for AGN using MaNGA data have used the [NII]- and [SII]-BPT only, in many cases in combination with the WHAN diagram. This biases the identification of AGN towards galaxies with solar to super-solar metallicities, missing a large fraction of low-metallicity galaxies hosting AGN (e.g., \citealt{2022ApJ...931...44P}). In addition, many searches have made use of integrated spectra within a (usually 3-arcsec radius) central aperture (e.g., \citealt{2017MNRAS.472.4382R}; \citealt{2018RMxAA..54..217S}; \citealt{2022MNRAS.514.3626C}) or the MaNGA field-of-view of the galaxy (e.g., \citealt{2022AJ....164..127C}). Those works using the spatially resolved information provided by the MaNGA IFU identify AGN using a certain AGN/LINER spaxel fraction and the distance between the AGN and star-formation demarcation line as an AGN metric (e.g., \citealt{2018MNRAS.474.1499W}; \citealt{2022ApJ...927...23C}) or a certain number of AGN/LINER spaxels classified as such by the [NII]- and [SII]-BPT (MDS+2020). This yields AGN fractions in dwarf galaxies of $\lesssim5$ \% (MDS+2020), much smaller than the $\sim$16\%-30\% obtained by \cite{2022ApJ...931...44P} when using integrated spectra but considering both the [NII]-, [SII]-, and [OI]-BPT in order not to miss AGN in low-metallicity galaxies. Yet, the bolometric luminosity of the AGN in the MDS+2020 dwarf galaxy sample was $\lesssim 10^{40}$ erg s$^{-1}$, the faintest among those found in previous optical, X-ray or radio searches, and the MaNGA data allowed to reveal a new type 1 AGN with log M$_\mathrm{BH} <10^{6}$ M$_{\odot}$. Given the power of IFU spectroscopy for revealing hidden and faint AGN in dwarf galaxies, in this paper we expand the work of MDS+2020 to the final MaNGA data release (DR17\footnote{\url{https://www.sdss.org/dr17/manga/}}) and perform a thorough identification of AGN using a pioneering spaxel by spaxel classification based on three spatially resolved emission line diagnostic diagrams ([NII]-, [SII]-, and [OI]-BPT; e.g., \citealt{2022ApJ...931...44P}) together with the WHAN diagram. As a result we obtain a sample of 664 AGN dwarf galaxies, the largest so far found in searches for AGN in dwarf galaxies, and which will be used as a benchmark for subsequent studies of the stellar population properties, AGN outflows, and merger activity in dwarf galaxies. 

% in a $\Lambda$-Cold Dark Matter (CDM) Universe in which dwarf galaxies are the smallest and most abundant blocks of structure formation. 
%also alleviate some of the discrepancies of the $\Lambda$-Cold Dark Matter (CDM) model such as the cusp versus core problem

The sample selection and analysis are described in Section~\ref{sample}, the results obtained are reported in Section~\ref{discussion}, conclusions are provided in Sect.~\ref{conclusions}. Throughout the paper we adopt a $\Lambda$-CDM cosmology with parameters $H_{0}=73$ km s$^{-1}$ Mpc$^{-1}$, $\Omega_{\Lambda}=0.73$ and $\Omega_{m}=0.27$. 

\section{Sample and analysis}
\label{sample}
The MaNGA DR17 is the final data release that contains all the observations and science data products for the complete survey, which includes over 10,000 galaxies at redshift z$<$ 0.15. MaNGA uses 17 IFUs, each composed by bundles of 19 to 127 fibers with a diameter of 2 arcsec, to cover a wavelength range of 3,600-10,300 \r{A} across 1.5 to 2.5 effective radii ($R_\mathrm{eff}$) at a spectral resolution R$\sim$ 2,000 and instrumental velocity dispersion of 70 km s$^{-1}$ (1 $\sigma$; \citealt{2021AJ....161...52L}).
The MaNGA data-analysis pipeline\footnote{\url{https://www.sdss.org/dr17/manga/manga-analysis-pipeline/}} (DAP) performs a stellar-continuum modeling and provides emission line measurements and 2D maps of measured properties. The size of the square spaxels in the maps are of 0.5 arcsec thanks to a dithering observational strategy (\citealt{2015AJ....150...19L}). In this paper we make use of the HYB10-MILESHC-MASTARHC2 maps {\footnote{The choice of HYB10-MILESHC-MASTARHC2 maps against other spectral fitting provided by DR17 is motivated by the more realistic H$\alpha$/H$\beta$ ratios obtained with that methodology (see Figure 5 in \citealt{2023MNRAS.518.4713B}.}, in which a Voronoi binning algorithm (\citealt{2003MNRAS.342..345C}) has been used to bin the spaxels to a signal-to-noise ratio (SNR) $\sim$10 for the kinematic analysis but the emission-line measurements have been performed on the individual spaxels. This provides greater spatial resolution for emission-line analysis, with a median Full Width Half Maximum (FWHM) of 2.5 arcsec, while avoiding the limitations of tying the analysis to the SNR of the broadband continuum. We also take from the MaNGA DAP the total star-formation rate (SFR) estimated from the Gaussian-fitted H$\alpha$ flux within the IFU field-of-view. The RA and DEC (from now on referred to as `optical center' of the galaxy), redshift ($z$) and $M_{*}$ are taken from the NASA-Sloan Atlas (NSA catalog) version v1\_0\_1, which is the one used by MaNGA for its targeting\footnote{\url{https://www.sdss.org/dr17/manga/manga-target-selection/nsa/}}. 
 
We draw our sample of dwarf galaxies as in MDS+2020: we consider only those galaxies with Petrosian and S\'ersic NSA stellar masses consistent within 0.5 dex (\citealt{2020ApJ...900...56S}) and then select dwarf galaxies as having $M_\mathrm{*} < 3 \times 10^{9}$ M$_{\odot}$ (similar to that of the Large Magellanic Cloud and typically considered in searches for AGN in dwarf galaxies, e.g., \citealt{2013ApJ...775..116R,2020ApJ...888...36R}; \citealt{2018MNRAS.478.2576M,2019MNRAS.488..685M}). To avoid biases towards galaxy type we do not apply any magnitude cut (e.g., \citealt{2013ApJ...775..116R}). Yet, all the dwarves have absolute r-band magnitude $M_\mathrm{r} > -20$ (see Fig.~\ref{ReffAbsMag}).
The stellar mass cut yields an initial sample of 3436 dwarf galaxies, of which 3,400 are unique galaxies\footnote{The DR17 sample contains a number of galaxies that have been observed more than once, either deliberately for data quality assessment purposes or due to early issues with targeting and/or plate drilling.} according to the DUPL\_GR parameter from the MaNGA Deep Learning Morphology catalogue (\citealt{2022MNRAS.509.4024D}, see also \citealt{2019MNRAS.483.2057F}; see Fig.~\ref{histNSAproperties}).

\begin{figure}
\centering
\includegraphics[width=0.49\textwidth]{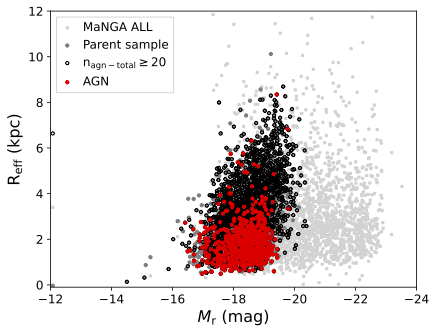}
\caption{Effective radius $R_\mathrm{eff}$ versus r-band NSA absolute magnitude corrected for extinction for the full MaNGA sample of $\sim$10,000 galaxies (grey squares), the parent sample of 3,400 unique dwarf galaxies (grey circles), the sample of 2,362 dwarf galaxies with strong indication of AGN ionisation (black circles), and the sample of 664 AGN candidates (red circles).}
\label{ReffAbsMag}
\end{figure}

\begin{figure}
\centering
\includegraphics[width=0.49\textwidth]{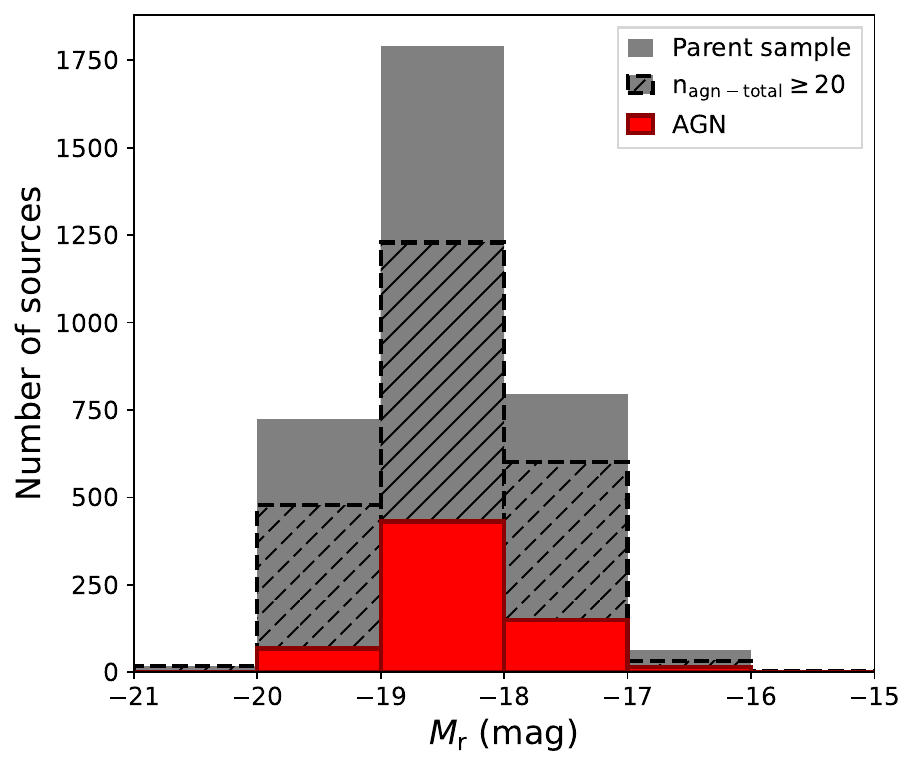}
\includegraphics[width=0.49\textwidth]{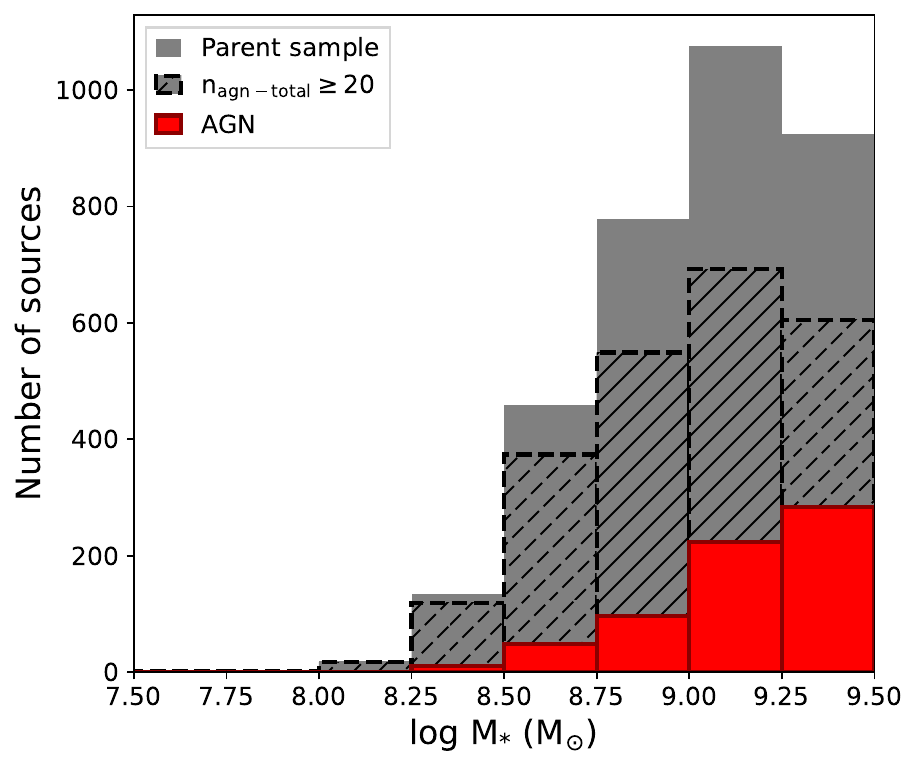}
\includegraphics[width=0.49\textwidth]{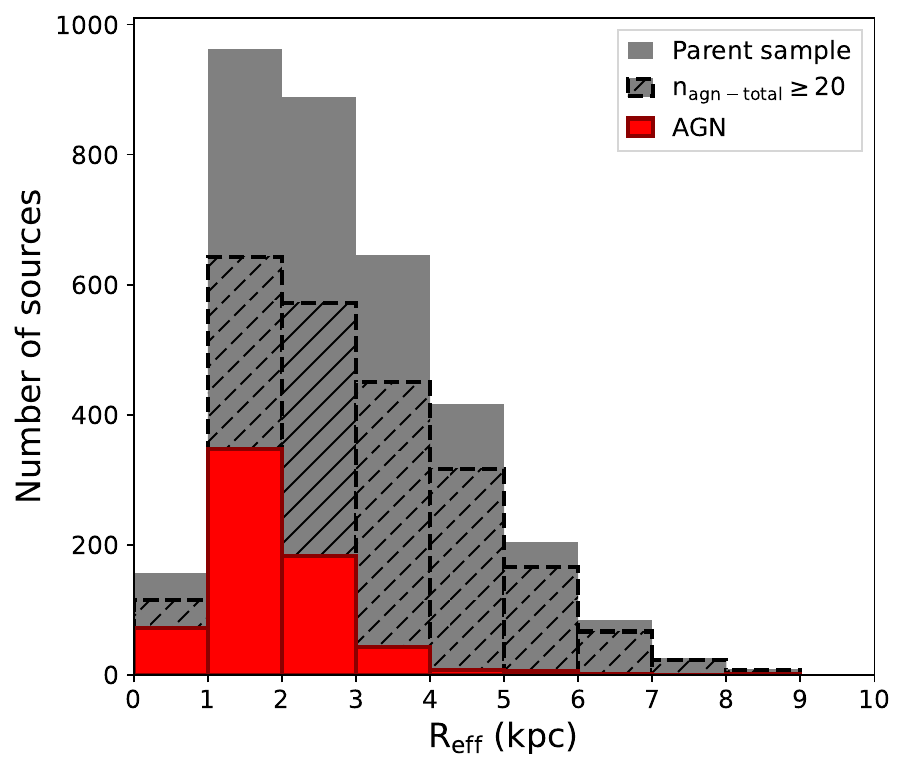}
\caption{Distribution of the r-band NSA absolute magnitude corrected for extinction (top), of the NSA Petrosian stellar masses that are consistent within 0.5 dex with the S\'ersic ones (middle), and of the effective radius $R_\mathrm{eff}$ (bottom) for the parent sample of 3,400 unique dwarf galaxies, the sample of 2,362 dwarf galaxies with strong indication of AGN ionisation, and the sample of 664 AGN candidates.}
\label{histNSAproperties}
\end{figure}

\subsection{Emission line diagnostic classification of individual spaxels}
\label{spaxclass}
To identify AGN among the sample of dwarf galaxies we use three optical emission line diagnostic diagrams: the [NII]-BPT, the [SII]-BPT, and the [OI]-BPT. We use the \cite{2001ApJ...556..121K,2006MNRAS.372..961K} and \cite{2003MNRAS.346.1055K} demarcation lines to differentiate between AGN, star-formation (SF), and Composite (mixture of AGN and SF) ionisation in the [NII]-BPT, and between Seyfert, SF, and LINER ionisation in the [SII]-BPT and [OI]-BPT. 
Hot old (post-AGB) stars can mimic LINER emission line ratios and are characterized by an H$\alpha$ equivalent width (EW) $<$ 3\AA. Hence, we additionally make use of the WHAN diagram (\citealt{2010MNRAS.403.1036C}) to differentiate between true AGN (EW(H$\alpha$) $>$ 3\AA\ and flux ratio log [NII]/H$\alpha >$ -0.4) and SF. 

Based on these diagnostics, each spaxel is classified as follows:
\begin{itemize}
\item AGN: if AGN or Composite in the [NII]-BPT and Seyfert in the [SII]- or [OI]-BPT. Color-coded as red in Figs.~\ref{BPT-AGN}-\ref{BPT-SF-AGN}.
\item AGN-WHAN: if AGN in the [NII]-BPT and SF or LINER in the [SII]- or [OI]-BPT.  Color-coded as red in Figs.~\ref{BPT-AGN}-\ref{BPT-SF-AGN}.
\item Composite: if Composite in the [NII]-BPT and SF or LINER in the [SII]- or [OI]-BPT. Color-coded as green in Figs.~\ref{BPT-AGN}-\ref{BPT-SF-AGN}.
\item SF-AGN: if SF in the [NII]-BPT and Seyfert in the [SII]- or [OI]-BPT. Color-coded as red in Figs.~\ref{BPT-AGN}-\ref{BPT-SF-AGN}.
\item LINER: if SF in the [NII]-BPT and i) LINER in the [SII]-BPT and SF in the [OI]-BPT or ii) SF in the [SII]-BPT and LINER in the [OI]-BPT. Color-coded as orange in Figs.~\ref{BPT-AGN}-\ref{BPT-SF-AGN}.
\item SF: SF in the [NII]-, [SII]- and [OI]-BPTs. Color-coded as blue in Figs.~\ref{BPT-AGN}-\ref{BPT-SF-AGN}.
\end{itemize}

We only consider those spaxels with SNR$\geq$3 for the emission lines used in each BPT (H$\alpha$, H$\beta$, [NII]$\lambda$6583, [SII]$\lambda$6718+[SII]$\lambda$6732, [OIII]$\lambda$5007, [OI]$\lambda$6300) and with an unmasked stellar velocity and stellar velocity dispersion (to ensure that continuum subtraction in that spaxel is correct). Examples of how the fits to these lines vary with SNR are shown in Figs.~\ref{BPT-AGN}-\ref{BPT-SF-AGN}. A discussion on the effects of the SNR threshold selection can be found in Appendix \ref{SNR}.

For each galaxy we then compute the total number of spaxels classified as either AGN, AGN-WHAN, or SF-AGN ($n_\mathrm{agn-total}$), as well as the total number of spaxels that are valid, this is, with SNR$\geq$3 and unmasked stellar velocity and stellar velocity dispersion ($n_\mathrm{valid}$). 

Applying the above criteria, 3,286 out of the 3,400 dwarf galaxies are found to have at least one AGN, AGN-WHAN or SF-AGN spaxel (this is, $n_\mathrm{agn-total} > 0$). 
%3322 out of the 3436 dwarf galaxies are found to have at least one AGN, AGN-WHAN or SF-AGN spaxel (this is, $n_\mathrm{total} > 0$). Thirty-six of these galaxies are found to be duplicated according to the DUPL\_GR parameter. After removing the duplicates we end up with 3286 dwarf galaxies (out of 3400) that have some indication of AGN ionisation (i.e. more than one spaxel). 

\begin{figure}
\centering
\includegraphics[width=0.49\textwidth]{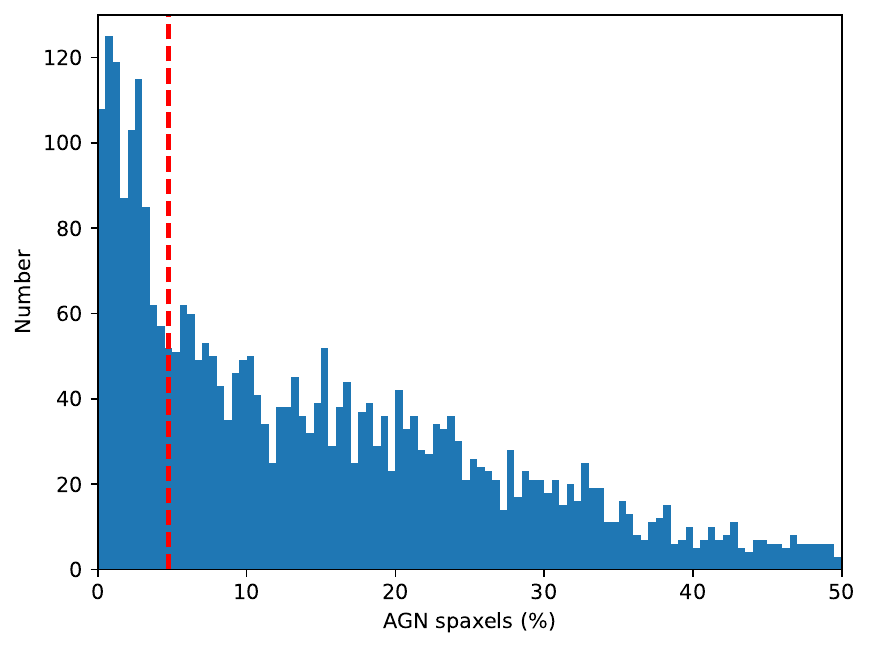}
\caption{Fraction of AGN spaxels (AGN + AGN-WHAN + SF-AGN) with respect to the total number of valid spaxels ($n_\mathrm{total}$/$n_\mathrm{valid}$, as defined in Sect. \ref{AGNclassification}) for the sample of 3,286 dwarf galaxies. The vertical dashed line marks the significant drop of the population at a fraction of 5\%.}
\label{Hist_AGNspaxel}
\end{figure}

\begin{figure}
\centering
\includegraphics[width=0.49\textwidth]{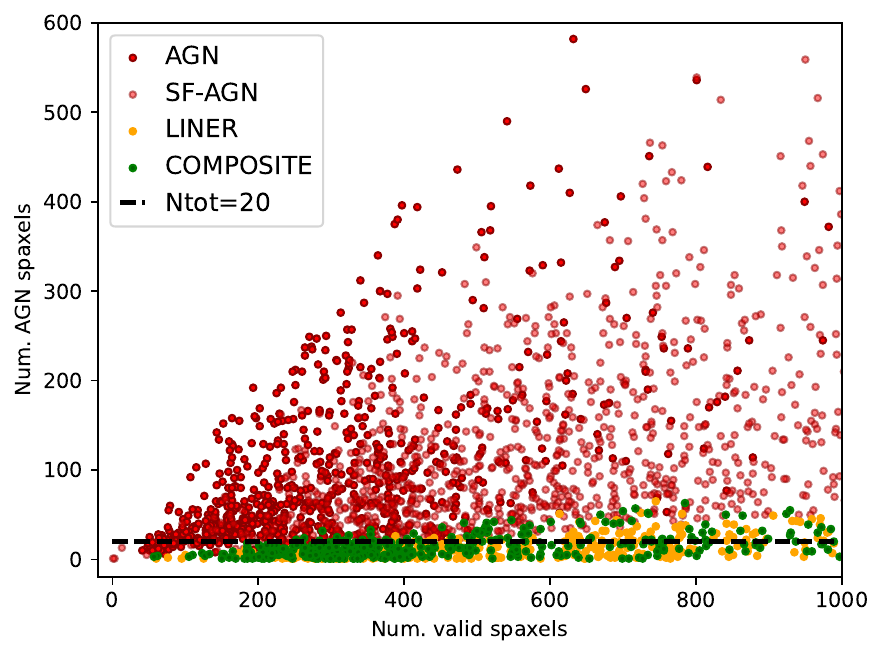}
\caption{Total number of AGN spaxels (AGN + AGN-WHAN + SF-AGN) versus number of valid spaxels ($n_\mathrm{agn-total}$ vs $n_\mathrm{valid}$) for the sample of 705 AGN, 1,203 SF-AGN, 445 Composite and 552 LINER dwarf galaxies selected based on the 5\% drop shown in Fig. \ref{Hist_AGNspaxel}. The horizontal dashed line marks the $n_\mathrm{agn-total}$ = 20, which we use as threshold to further reduce our sample.}
\label{AGNspaxels_validspaxel}
\end{figure}

\subsection{Emission line diagnostic classification of each galaxy}
\label{AGNclassification}
After performing a spaxel by spaxel classification, we plot a histogram of the fraction of AGN + AGN-WHAN + SF-AGN spaxels that are valid ($n_\mathrm{agn-total}$/$n_\mathrm{valid}$; see Figure \ref{Hist_AGNspaxel}) for the sample of 3,286 dwarf galaxies. We then select as threshold the percentage at which there is a significant drop of the main population. We chose a percentage of 5\%, and perform a final classification of each galaxy as follows:
\begin{itemize}
\item AGN: if the ratio of AGN spaxels to valid spaxels ($n_\mathrm{agn}$/$n_\mathrm{valid}$) $\geq$5\%.
\item AGN-WHAN: if the ratio of AGN-WHAN spaxels to valid spaxels ($n_\mathrm{agn-whan}$/$n_\mathrm{valid}$) $\geq$5\% and the galaxy is not AGN. WHAN diagram to be applied.
\item SF-AGN: if the ratio of SF-AGN spaxels to valid spaxels ($n_\mathrm{sf-agn}$/$n_\mathrm{valid}$) $\geq$5\% and the galaxy is not AGN nor AGN-WHAN.
\item Composite: if the ratio of Composite spaxels to valid spaxels ($n_\mathrm{composite}$/$n_\mathrm{valid}$) $\geq$5\% and the galaxy is not AGN nor AGN-WHAN nor SF-AGN.
\item LINER: if the ratio of LINER spaxels to valid spaxels ($n_\mathrm{liner}$/$n_\mathrm{valid}$) $\geq$5\% and the galaxy is not AGN nor AGN-WHAN nor SF-AGN nor Composite.
\end{itemize}

This results in 698 AGN, 337 AGN-WHAN, 1,203 SF-AGN, 445 Composite, and 552 LINER dwarf galaxies. Applying the WHAN diagram to the 337 AGN-WHAN we obtain that seven of the galaxies are classified as AGN. Including these seven AGN-WHAN that are AGN according to the WHAN diagram, we have that out of 3,286 dwarves there are 705 AGN. The number of $n_\mathrm{agn-total}$ vs. $n_\mathrm{valid}$ spaxels for these 705 AGN, 1203 SF-AGN, 445 Composite and 552 LINER dwarf galaxies is plotted in Fig. \ref{AGNspaxels_validspaxel}.
At $n_\mathrm{agn-total}$ = 20 a clear separation can be observed between AGN and LINER/Composite galaxies. To be conservative and define an AGN sample as robust as possible, we thus apply a further cut to select only sources with $n_\mathrm{agn-total} \geq$ 20. This removes 323 Composite and 379 LINER galaxies from our sample, but only 27 SF-AGN and 41 AGN, none of which are AGN based on other diagnostics (integrated SDSS emission line classification, radio emission, X-ray emission, or infrared emission; see Sects. \ref{SDSS} and \ref{multiwavelength}). We find that this cut is the most balanced one between number of AGN and Composite/LINER galaxies excluded. For instance, using a threshold of $n_\mathrm{agn-total}$ = 10 we would only miss 2 AGN but include still hundreds of Composite and LINER dwarf galaxies. Increasing the threshold to $n_\mathrm{agn-total}$ = 50, we would miss 1,720 sources, of which 254 are AGN. Five of these AGN are also classified as such based on radio and infrared diagnostics. Therefore, our final sample is selected by applying a cut at $n_\mathrm{agn-total} \geq$ 20. This reduces the sample of 3,286 dwarf galaxies with some indication of AGN ionisation (i.e. $n_\mathrm{agn-total} >$ 0) to 2,362 galaxies with strong indication of AGN ionisation ($n_\mathrm{agn-total} \geq$ 20), of which 664 are AGN, 1,176 SF-AGN, 122 Composite, and 173 LINERs (see Fig. \ref{flowchart}). The 664 AGN constitute our primary sample of AGN dwarf galaxies. Examples of the spatially-resolved spaxel by spaxel emission line diagnostic classification for an AGN and a SF-AGN dwarf galaxy are shown in Figs.~\ref{BPT-AGN}-\ref{BPT-SF-AGN}.

\begin{figure}
\centering
\includegraphics[width=0.49\textwidth]{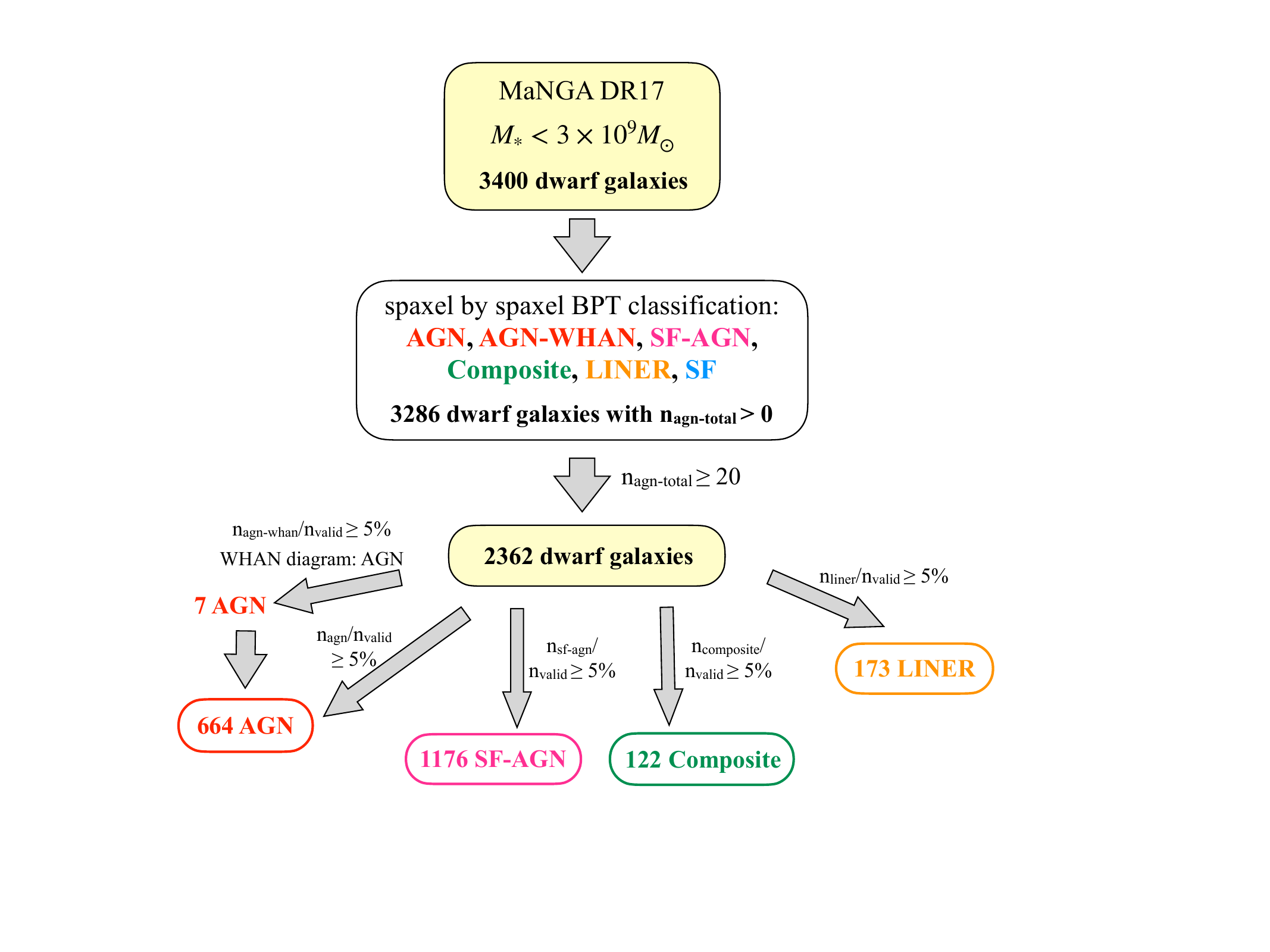}
\caption{Flow chart summarizing the selection criteria.}
\label{flowchart}
\end{figure}

\section{Results and Discussion}
\label{discussion}
The application of the above emission line diagnostic criteria to the parent sample of 3,400 dwarf galaxies results in 2,362 galaxies with strong AGN ionisation features ($n_\mathrm{agn-total} \geq$ 20), among which 664 AGN, 1,176 SF-AGN, 122 Composite, and 173 LINERs are identified. The properties of ten of these sources are provided in Table \ref{AGNcatalog}, while the full catalogue can be found in the online version of the paper. To investigate whether these sources are identified as AGN by single-fiber integrated emission line diagnostics, we use their SDSS emission line ratios and crossmatch with catalogues of type 1 AGN. We also check whether they are included in previous MaNGA samples of AGN and whether they have counterparts at other wavelengths (infrared, X-rays, radio) indicating an AGN nature. 

\begin{figure*}
\centering
\includegraphics[width=\textwidth]{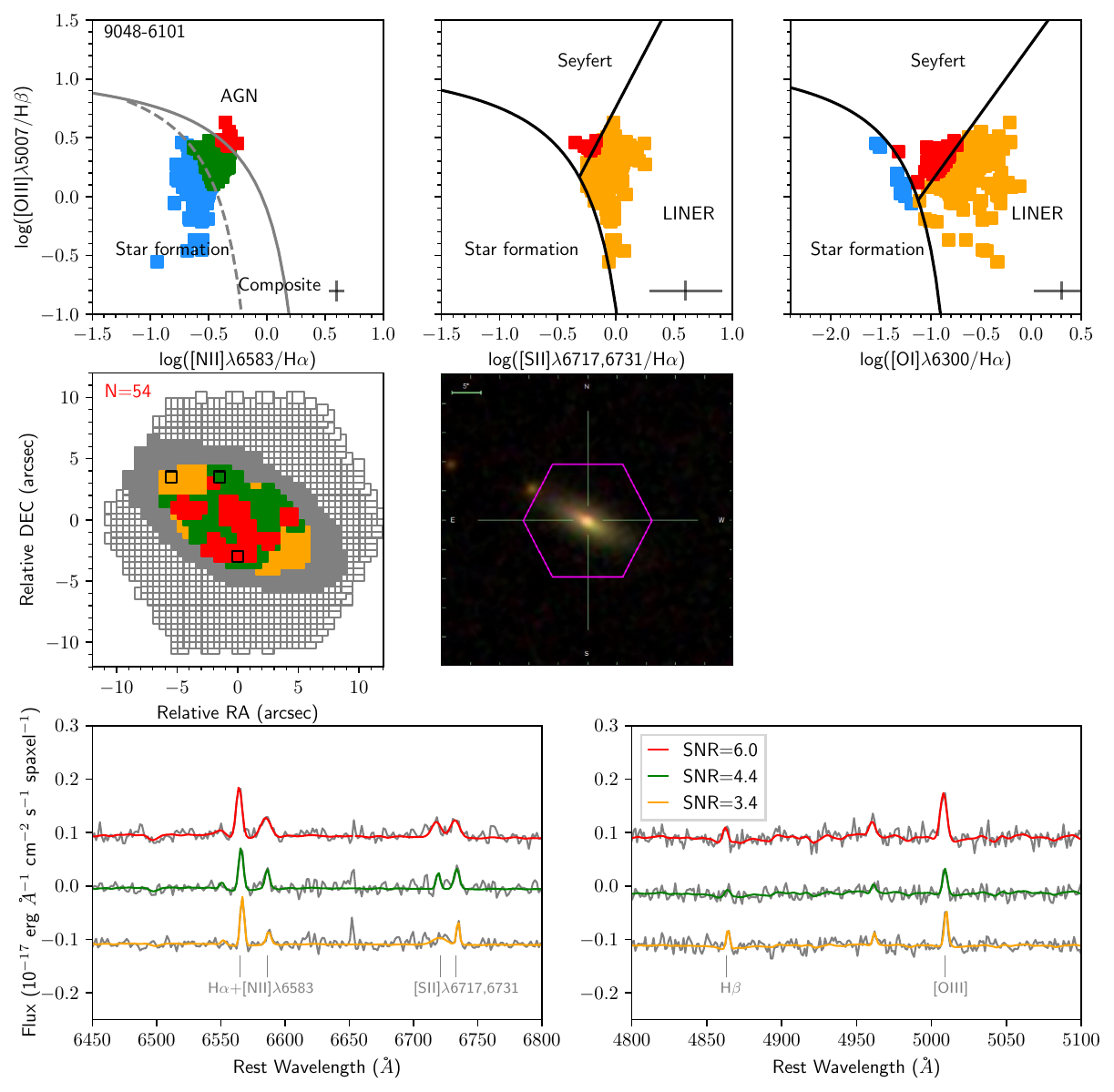}
\caption{MaNGA analysis for one of the AGN dwarf galaxies. \textrm{Top row}: Location of each MaNGA spaxel on the [NII]-BPT (left), [SII]-BPT (middle), and [OI]-BPT (right) used to distinguish between ionisation by AGN (red spaxels), star-formation (blue spaxels), Composite (green spaxels), and LINER (orange spaxels). The grey crosses indicate the median of the uncertainty on the flux ratios. \textrm{Middle left}: Spatial distribution of the BPT-classified spaxels (color-coded as described in Sect. \ref{spaxclass}). Empty squares mark the IFU coverage, grey squares those spaxels with continuum SNR $>$ 1. The `N' shows the number of AGN spaxels used in the analysis and stacking (Sect. \ref{BHmass}). The squares marked in black correspond to the three spaxels whose spectra are shown in the bottom panels. \textrm{Middle right}: SDSS composite image. The pink hexagon shows the IFU coverage. \textrm{Bottom}: Spectrum (grey line) of the spaxels marked in black in the Middle left panel. The full MaNGA DAP model is shown in red (for the AGN spaxel), in green (for the Composite spaxel), and in orange (for the LINER spaxel). The spectra have been offset for visualization purposes. The SNR of each of these spaxels is shown in the legend.}
\label{BPT-AGN}
\end{figure*}

\begin{figure*}
\centering
\includegraphics[width=\textwidth]{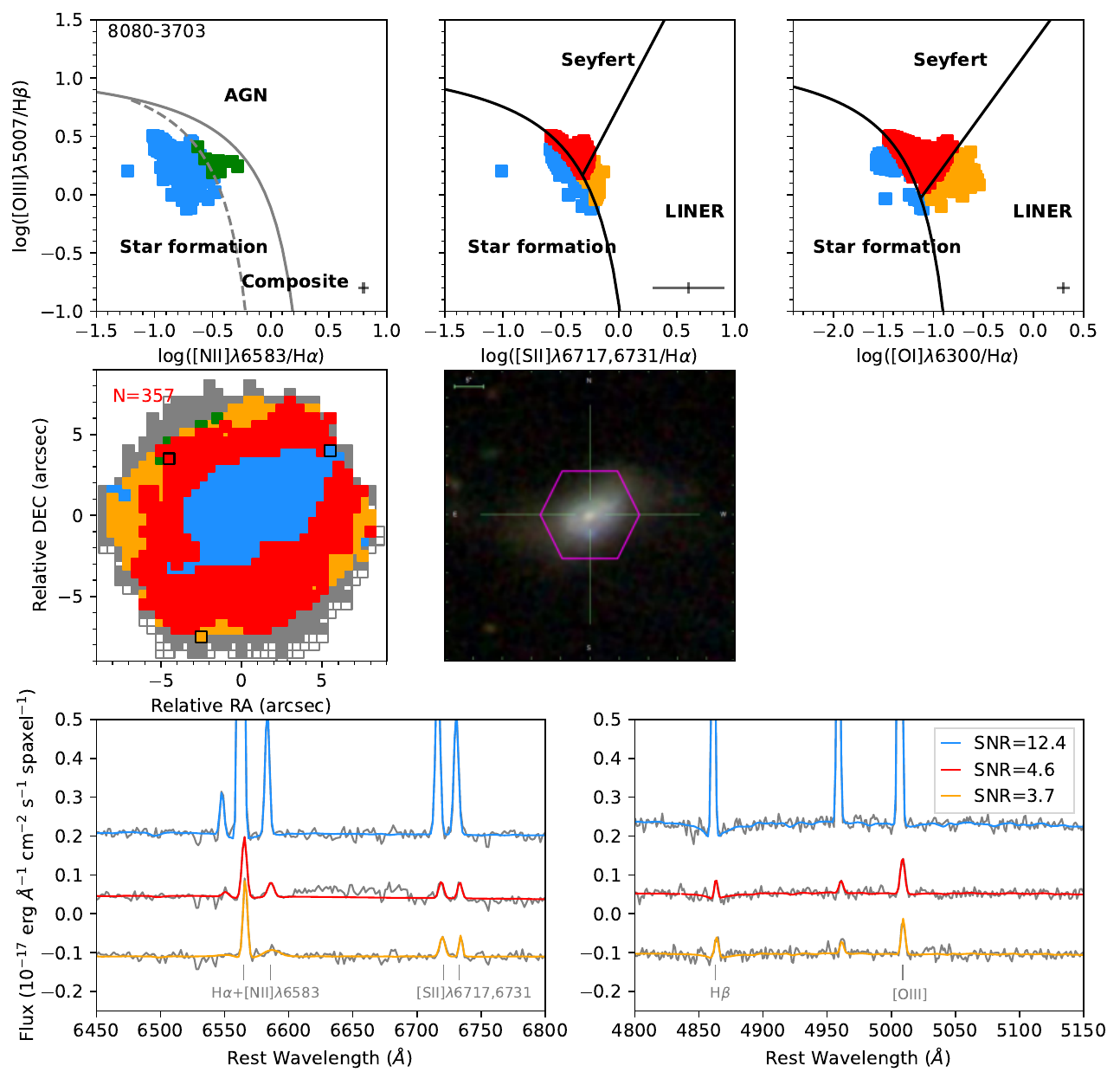}
\caption{Same caption as in Fig.~\ref{BPT-AGN} but for a SF-AGN dwarf galaxy. The blue line in the spectra shown in the bottom panels corresponds to the full MaNGA DAP model of the SF spaxel marked in black in the Middle left panel.}
\label{BPT-SF-AGN}
\end{figure*}

\begin{table*}
\footnotesize{}
\label{AGNcatalog}
\caption{Exemplary ten rows of the sample of 2,362 dwarf galaxies with strong AGN ionisation features ($n_\mathrm{agn-total} \geq$ 20).}
\begin{tabular}{{lcccccccl}}
\hline
\hline 
MaNGA & MaNGA & RA (J2000) & DEC (J2000) & $z$ &  log $M_\mathrm{*}$ & SFR & Type & Other\\
plateifu & ID & (deg) & (deg) &  & (M$_{\odot}$) & (M$_{\odot}$ yr$^{-1}$) & & \\
(1) & (2)   & (3)   & (4)   & (5)   & (6)   & (7)   &  (8) & (9)\\
\hline
 9886-3701 & 1-594067 & 237.785510 & 25.722870 &  0.022 &          8.3 & 0.002264 &     AGN &       -- \\
 9886-3702 & 1-317589 & 238.020919 & 26.074462 &  0.021 &          8.5 & 0.035987 &  SF-AGN &       -- \\
 9886-6102 & 1-317060 & 237.961673 & 25.937835 &  0.022 &          8.4 & 0.021797 &  SF-AGN &       -- \\
 9886-9101 & 1-375801 & 237.371264 & 24.876893 &  0.024 &          8.9 & 0.046428 &  SF-AGN &       -- \\
 9889-1902 & 1-440292 & 234.858600 & 24.943569 &  0.023 &          9.2 & 0.188022 &     AGN &    RADIO \\
 9889-6102 & 1-440301 & 234.789840 & 24.830893 &  0.016 &          8.7 & 0.045221 &  SF-AGN &       -- \\
9890-12705 & 1-316607 & 233.520461 & 29.906882 &  0.037 &          8.9 & 0.072680 &  SF-AGN &       -- \\
9891-12703 & 1-593748 & 229.324480 & 29.400540 &  0.018 &          8.7 & 0.008508 &  SF-AGN &       -- \\
 9891-1901 & 1-593648 & 227.177572 & 28.171156 &  0.026 &          9.4 & 0.218527 &     AGN &       -- \\
 9892-9102 & 1-295128 & 248.989901 & 24.104075 &  0.037 &          8.8 & 0.039987 &  SF-AGN &       -- \\

\hline
\hline
\end{tabular}
\smallskip\newline\small {\bf Column designation:}
(1) MaNGA plateifu; (2) MaNGA ID; (3,4) RA, DEC coordinates of the optical center of the galaxy or IFU center; (5) galaxy redshift; (6) galaxy stellar mass; (7) galaxy star-formation rate within the IFU FoV as provided by the MaNGA DAP (i.e. based on the H${\alpha}$ flux and uncorrected for extinction); (8) MaNGA BPT classification; (9) AGN in other diagnostics (radio, X-rays, mid-infrared --MIR--, or coronal lines --CL--). The full catalogue can be found in the online version of the paper.
\end{table*}

\subsection{SDSS emission line diagnostics}
\label{SDSS}
We crossmatch the sample of 2,362 dwarf galaxies with the emission-line catalogue of the Portsmouth Group\footnote{\url{https://www.sdss.org/dr12/spectro/galaxy_portsmouth/}}, which includes emission-line measurements of the SDSS DR12, and select only those sources with SNR$\geq$3 in the emission lines used in each BPT. This results in 1,575 dwarf galaxies with reliable SDSS emission line measurements. 
Among these 1,575 dwarf galaxies, there are 284 of our MaNGA AGN. However, of these 284 AGN only 52 are classified as AGN when using the SDSS emission line measurements and applying our same criteria as for the MaNGA sample (i.e. AGN or LINER in the [NII]-BPT, AGN in the [SII]-BPT, or AGN in the [OI]-BPT). 
Because type 1 AGN do not have emission-line measurements in the Portsmouth Group catalogue, we also crossmatch our sample of 2,362 dwarf galaxies with a) the type 1 AGN catalogue of \cite{2019ApJS..243...21L}, which includes 14,584 broad-line AGN at z$<$0.35 drawn from SDSS DR7, and b) the 305 IMBH candidates with broad H$\alpha$ emission of \cite{2018ApJ...863....1C}, also drawn from SDSS DR7 using an automated data mining workflow. This results in five and three, respectively, broad-line AGN (see Sect.~\ref{BHmass}), all of which are classified as AGN when using MaNGA data. Of the five broad-line AGN from \cite{2019ApJS..243...21L}, two are already included in the sample of 52 AGN identified by the BPT using the SDSS emission lines. Hence the total number of AGN classified as such when using SDSS emission line measurements is of 58 (out of the 284 that are AGN in MaNGA). The use of integrated SDSS spectra thus implies missing $\sim$80\% of our sample of AGN.
%((284-52-6)/284*100)=79.57percent 

The numbers are even more dramatic for the SF-AGN. There are 1,041 of our MaNGA SF-AGN among the 1,575 dwarves with reliable Portsmouth emission line measurements. Of these, only seven are classified as SF-AGN (i.e. SF in the [NII]-BPT and AGN in the [SII]- or [OI]-BPT) when using the SDSS fluxes. Nearly all of the SF-AGN MaNGA dwarf galaxies would thus be missed when using SDSS spectroscopy to search for SF-AGN (see also \citealt{2022ApJ...931...44P}).
%((1041-7)/1041*100$\gtrsim$99\%)

We also check how many of the dwarf galaxies are identified as AGN based on the SDSS emission line fluxes derived by \cite{2022AJ....163..224H}, who apply the spectroscopic fitting package GELATO to 416,288 SDSS galaxies. Their sample includes 1,322 of our 2,362 dwarf galaxies, and all have SNR$\geq$3 in the emission lines used in the [NII]-, [SII]-, and [OI]-BPTs. There are 308 of our AGN among the 1,322 dwarves, of which only 43 are classified as AGN  based on the SDSS  emission line fluxes of \cite{2022AJ....163..224H}. This implies that 86\% of the AGN are not identified as AGN when using integrated spectra, similarly to what was found when using the Portsmouth Group emission line measurements. In the case of SF-AGN, there are 736 of our SF-AGN among the 1,322 dwarves from \cite{2022AJ....163..224H}. However, none of them are identified as such when using the SDSS measurements from \cite{2022AJ....163..224H}. Hence all of the SF-AGN MaNGA dwarf galaxies with Hviding's emission lines fluxes are missed when using SDSS integrated spectroscopy.
%((308-43)/308)=86percent

\subsection{AGN diagnostics at other wavelengths}
\label{multiwavelength}
To further investigate the AGN nature of the sample of 2,362 dwarf galaxies we look for counterparts in the infrared, radio, and X-ray bands. 
All the galaxies are found to have a Wide-field Infrared Survey Explorer (WISE; \citealt{2010AJ....140.1868W}) mid-infrared counterpart within 10 arcsec; however, only 15 of them qualify as AGN according to the \cite{2011ApJ...735..112J} or \cite{2012ApJ...753...30S} criteria. Using the theoretical AGN cut of \cite{2018ApJ...858...38S}, we find that six additional sources qualify as AGN. We also apply the most recent and stringent cuts from \cite{2022AJ....163..224H}, finding 13 AGN already classified as such by the \cite{2011ApJ...735..112J}, \cite{2012ApJ...753...30S}, or \cite{2018ApJ...858...38S} criteria. Of the in total 21 (15+6) AGN dwarf galaxies according to WISE, 17 are classified as AGN, three as SF-AGN, and one as Composite by our emission line diagnostic classification. (Sect.~\ref{AGNclassification}).  

In the radio, we find that out of the 2,362 dwarf galaxies, 55 have a Faint Images of the Radio Sky at Twenty Centimeters (FIRST; \citealt{1995ApJ...450..559B}) counterpart and 22 a 1.4GHz NRAO VLA Sky Survey (NVSS; \citealt{1998AJ....115.1693C}) counterpart within 5 arcsec. Of the 22 NVSS counterparts, 17 sources have also a FIRST counterpart, so for these we consider the FIRST radio fluxes rather than the NVSS fluxes in the radio analysis. The total number of FIRST/NVSS radio counterparts (excluding duplicates) is thus of 60. The integrated radio emission ranges from 0.7 to 31.6 mJy and is resolved in 35 ($\sim$53\%) of the sources. We use the SFR within the IFU FoV and the correlations from \citeauthor{2019MNRAS.484..543F} (2019, see \citealt{2019MNRAS.488..685M}) to compute the expected (thermal and non-thermal) contribution from star formation to the radio emission. The SFR is corrected for extinction using the \cite{2000ApJ...533..682C} extinction law and the total integrated flux of the H$\alpha$ and H$\beta$ emission lines within the full MaNGA FoV to compute the Balmer decrement (H$\alpha$/H$\beta$) and color excess [E(B-V)]. To derive the maximum luminosity expected from a supernova remnant and supernova and the total luminosity expected from a population of supernova remnants and supernovae we use equations 5 and 6, respectively, in \cite{2020ApJ...888...36R}.
We find that in $\sim$27\% (16 out of 60) of the detections the radio emission is $\geq3\sigma$ above that expected from stellar processes and thus consistent with AGN. Among the 16 radio AGN, $\sim$69\% are AGN and $\sim$19\% SF-AGNs according to our MaNGA emission line diagnostic classification. Three of these sources are also classified as AGN according to the WISE diagnostics.
%11 robust AGN, 3 SF-AGN, 0 Composite, and 0 LINERs 

In the X-rays, we search for counterparts in the Chandra Source Catalog (\citealt{2010ApJS..189...37E}) Version 2 (CSC 2.0), whose sky coverage totals $\sim$560 deg$^2$. Out of 2,362 dwarves, 177 have been observed by \textit{Chandra} and are within the CSC footprint. Of these, there are 162 non-detections and 15 detections. For the detections, we get from the Master Source Catalog\footnote{\url{https://cxc.harvard.edu/csc/columns/master.html}} the aperture-corrected net energy flux inferred from the source region aperture, averaged over all contributing observations, in the ACIS broad (0.5-7.0 keV) energy band. The corresponding X-ray luminosities range from $L_\textrm{0.5-7 keV}$ = 7.2 $\times$ 10$^{38}$ erg s$^{-1}$ to 4.5 $\times$ 10$^{42}$ erg s$^{-1}$. To investigate the origin of this X-ray emission we derive the 2-10 keV X-ray luminosity expected from X-ray binaries ($L_\textrm{XRB}$) using the relation between $L_\textrm{XRB}$,  $M_\mathrm{*}$, and SFR from \cite{2010ApJ...724..559L}. We convert the CSC 0.5-7 keV to the 2-10 keV band applying a conversion factor of 7.207 $\times$ 10$^{-16}$ erg cm$^{-2}$ s$^{-1}$ for an unabsorbed flux of 10$^{-15}$ erg cm$^{-2}$ s$^{-1}$, found with PIMMS\footnote{\url{https://cxc.harvard.edu/toolkit/pimms.jsp}} assuming gamma=1.8. Ten out of the 15 CSC detections are found to have X-ray luminosities more than 3$\sigma$ above that expected from X-ray binaries and are thus consistent with being X-ray AGN. 

We also look for X-ray counterparts in the fourth XMM-Newton serendipitous source (4XMM-DR11) catalogue (\citealt{2020A&A...641A.136W}). Using a search radius of 5 arcsec, we find 18 counterparts with flag=0 or 1 (i.e. no spurious detection) among the 2,362 dwarf galaxies. Their 0.2-12 keV luminosities range from $L_\textrm{0.2-12 keV}$ = 8.0 $\times$ 10$^{39}$ erg s$^{-1}$ to 3.9 $\times$ 10$^{42}$ erg s$^{-1}$, which we convert to 2-10 keV using PIMMS. The 2-10 keV luminosities are more then 3$\sigma$ above that expected from X-ray binaries for 17 out of the 18 sources, indicating an AGN origin of the X-ray emission. Six of the XMM AGN have also a CSC counterpart consistent with being X-ray AGN. 

One additional X-ray AGN is found among the 2,362 dwarf galaxies when crossmatching with the eROSITA Final Equatorial Depth Survey (eFEDS) main catalogue (\citealt{2022A&A...661A...1B}) using a radius of 3 arcsec, which is the nominal positional accuracy of eROSITA. The eFEDS catalog includes 27,910 X-ray sources detected in the 0.2-2.3 keV energy range. The 2.3-5 keV band luminosity of the source is 3.4 $\times$ 10$^{40}$ erg s$^{-1}$, which we convert to 2-10 keV using PIMMS. The resulting 2-10 keV luminosity is more than 5 times higher than that estimated for X-ray binaries using \cite{2010ApJ...724..559L}. 

In total, 26 out of the 2,362 dwarf galaxies are found to have X-ray emission, 88\% of which are consistent with AGN. Using our emission line diagnostics (Sect.~\ref{AGNclassification}), $\sim$48\% of these sources qualify as AGN, $\sim$17\% as SF-AGN, $\sim$17\% as Composite, and $\sim$9\% as LINER. Of these, two sources are also classified as AGN based on WISE diagnostics. 
%Of 23 Xray AGN: 11 of these sources qualify as robust AGN, 4 as SF-AGN, 4 as Composite, and 2 as LINER. The remaining 9\% correspond to a Retired galaxy and a SF one.

The detection of coronal lines has been also often used to infer the presence of AGN in dwarf galaxies, both in the infrared (e.g., \citealt{2007ApJ...663L...9S}; \citealt{2018ApJ...861..142C,2020ApJ...895..147C}) and, more recently, also in the optical (e.g., \citealt{2021ApJ...922..155M}; \citealt{2022ApJ...937....7S}) regimes. \cite{2023ApJ...945..127N} have recently performed a search for coronal lines ([NeV] $\lambda$3347, $\lambda$3427, [FeVII] $\lambda$3586, $\lambda$3760, $\lambda$6086, and [FeX] $\lambda$6374) to the latest MaNGA's DR17, finding 71 galaxies with emission from at least one of the coronal lines. Six of these galaxies are included in our sample: one as AGN, four as SF-AGN, and one LINER. None of them is classified as AGN according to any of the other multiwavelength diagnostics.

Therefore, in total, out of the 2,362 dwarf galaxies $\sim$3\% (61 sources) are classified as AGN according to either the radio, X-ray or mid-infrared emission or the detection of coronal lines. Of these, 35 are classified as AGN using our emission line diagnostics, 14 as SF-AGN, 5 as Composite and 3 LINER\footnote{The remaining sources are classified as Quenched or Retired by the WHAN diagram}. Cutouts from the Legacy Survey DR9 are shown in Fig.~\ref{cutouts} for a representative sample of 15 of the AGN with a counterpart at X-ray, radio, or mid-infrared wavelengths.
Follow-up X-ray and radio campaigns should be able to increase the statistics.  

\begin{figure*}
\centering
\includegraphics[width=\textwidth]{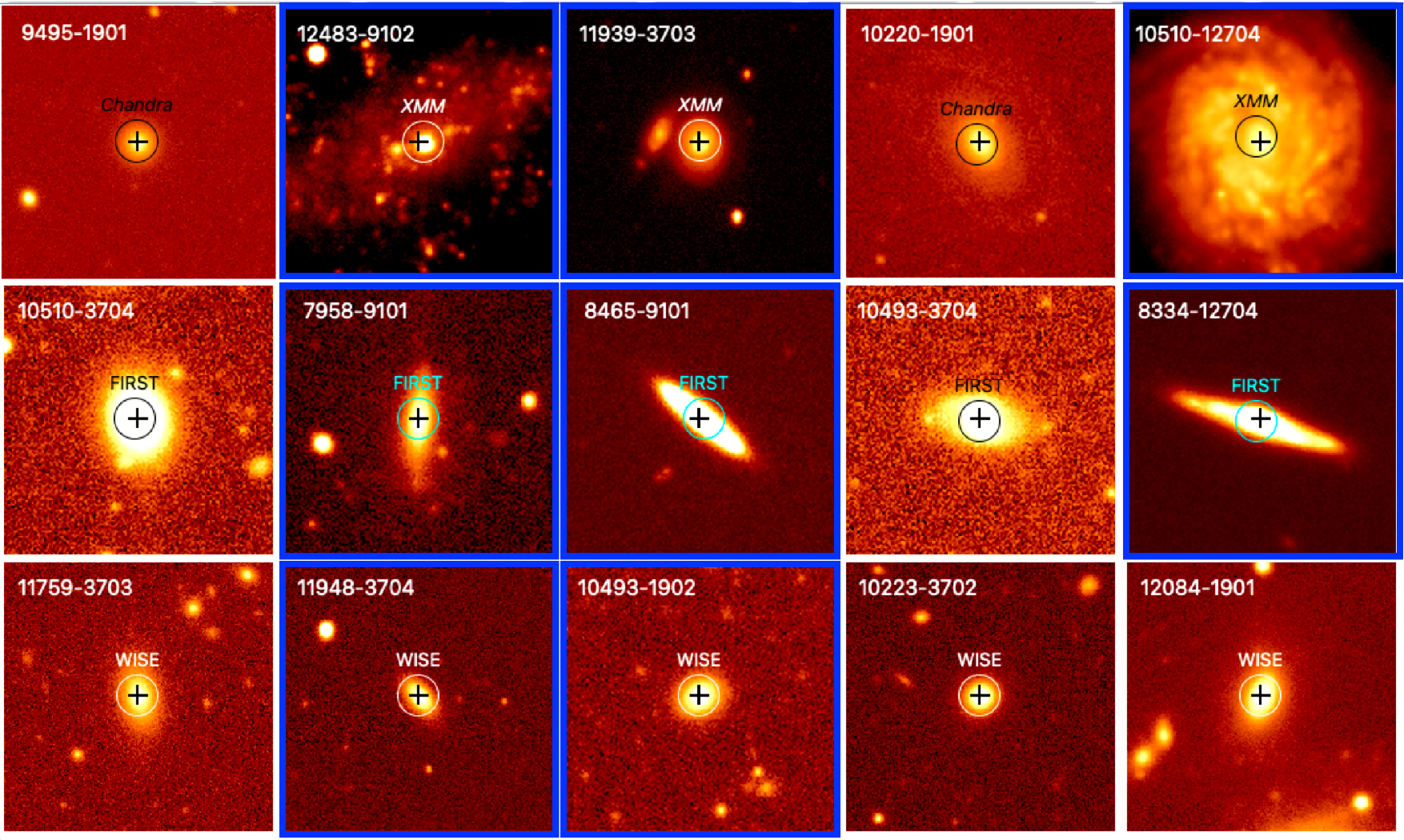}
\caption{Legacy Survey DR9 cutouts of a representative sample of 15 of the AGN classified as such based on our MaNGA analysis. The black crosses mark the MaNGA optical center. The circles, of 5 arcsec radius, mark the position of the \textit{Chandra} or \textit{XMM-Newton} counterparts (first row), FIRST radio counterparts (second row) and WISE mid-infrared counterparts (bottom row). Sources in each row are shown in increasing stellar mass (from left to right). Blue squares mark those galaxies classified as late-type (T-Type $>$ 0), while the remaining sources are early-type (T-Type $<$ 0). }
\label{cutouts}
\end{figure*}

\subsection{Off-nuclear AGN, obscuration, and AGN dilution}
\label{offnuclearAGN}
Most ($\sim$80\%) of the AGN and nearly all the SF-AGN in the sample of 2,362 dwarf galaxies are not identified as such based on SDSS emission line diagnostics (see Sect. \ref{SDSS}). For some of them this might be due to star formation dominating the optical emission, specially in the central parts sampled by single-aperture spectroscopy, as inferred from the finding of AGN signatures in the radio and X-ray regimes (see Sect. \ref{multiwavelength}). 
To further investigate whether the sources are off-nuclear AGN we derive the offset between the optical center of the galaxy and the median position of the AGN + AGN-WHAN + SF-AGN + LINER spaxels (from now on `optical offset'). The distribution of such offsets for the different sub-samples is shown in Fig.~\ref{offset}, top panel. The AGN sub-sample is the one with the major number of sources within 3 arcsec (250 out of 664, i.e. $\sim$38\%) while only 2\% of the SF-AGN have offsets within that radius, in agreement with the percentages of AGN and of SF-AGN identified with the SDSS (central spectroscopic fiber of 3 arcsec)\footnote{We note there is a slight offset, of median value 0.13 arcsec, between the MaNGA optical center and the SDSS spectroscopic fiber position; however this does not impact our results.}. The offset of the SF-AGN, Composite and LINER sub-samples is indeed found to be larger than that of the AGN, with median values of 6.1 arcsec $\pm$ 2.2 arcsec for the SF-AGN, 7.1 arcsec $\pm$ 2.4 arcsec for the Composite, 8.1 arcsec $\pm$ 2.2 arcsec for the LINERs, and of 3.5 arcsec $\pm$ 1.6 arcsec for the AGN. Applying a two-sample Kolmogorov-Smirnov test between the AGN and the SF-AGN/Composite/LINER sub-samples yields p-values smaller than 0.05 in all cases, rejecting the null hypothesis that the data are drawn from the same distribution when using a confidence level of 95\%. The SF-AGN, Composite and LINER sources are thus more likely to be either nuclear AGN obscured by star formation, off-nuclear AGN wandering in the dwarf galaxy, or switched-off AGN for which we detect the relics of the last burst of ionisation (i.e. MDS+2020). 

To distinguish between these scenarios we measure the Balmer decrement (H$\alpha$/H$\beta$) within the central 3 arcsec of each galaxy (by taking the median value of all the spaxels within that radius) and then derive the color excess E(B-V) considering the \cite{2000ApJ...533..682C} dust attenuation curves. Despite the differences observed in terms of offsets and SDSS identification among the AGN, SF-AGN, Composite and LINER sub-samples, no significant differences are obtained in terms of E(B-V), with median values of 0.2 $\pm$ 0.2 for the AGN, 0.1 $\pm$ 0.1 for the SF-AGN, 0.3 $\pm$ 0.1 for the Composite, and 0.2 $\pm$ 0.1 for the LINERs (see Fig.~\ref{offset}, bottom panel). No differences in the color excess are found either when dividing the whole sample of 2,362 dwarf galaxies into `nuclear AGN' (i.e. those with offsets $\leq$ 3 arcsec) and off-nuclear AGN (i.e. with offsets $>$ 3 arcsec), suggesting that obscuration is not the major driver of the AGN offset. 

\begin{figure}
\centering
\includegraphics[width=0.49\textwidth]{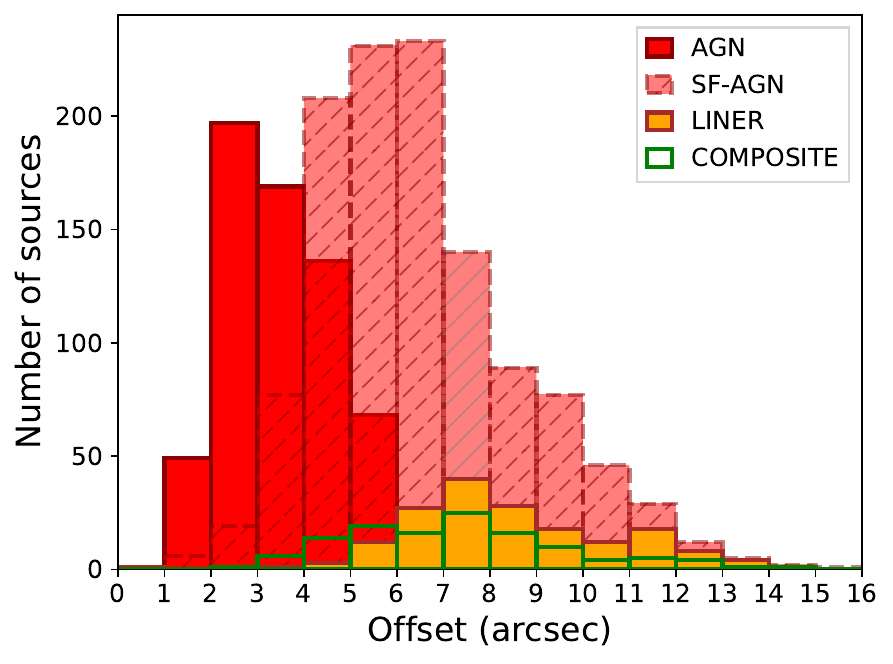}
\includegraphics[width=0.49\textwidth]{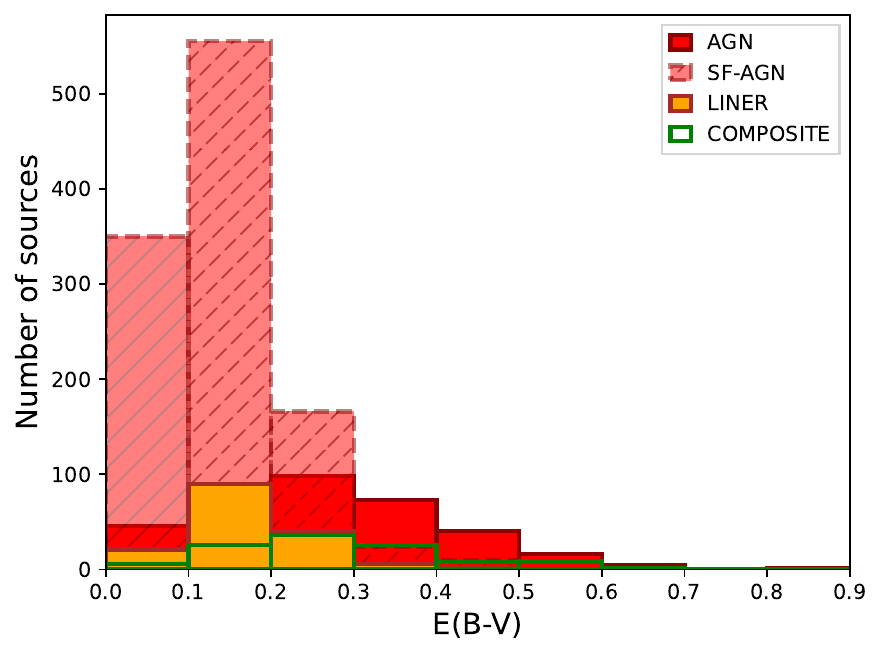}
\caption{Top: Distribution of the offset between the optical center of the galaxy and the median position of the AGN + AGN-WHAN + SF-AGN + LINER spaxels for the different sub-samples. Bottom: Distribution of the color excess within the central 3 arcsec for the different sub-samples. }
\label{offset}
\end{figure}

We next check whether dilution by star formation is responsible for the different offsets. We derive the SFR within the central 3 arcsec from the median of the H$\alpha$ flux in that region, using the \cite{2012ARA&A..50..531K} correlation (log SFR = log L$_\mathrm{H\alpha}$ - 41.27) as used by the MaNGA DAP to compute global SFRs, and correct if for extinction applying the \cite{2000ApJ...533..682C} dust attenuation curves. The distribution of SFRs within 3 arcsec for the AGN, SF-AGN, Composite and LINER sub-samples is shown in Fig.~\ref{SFR3arcsec_nuc_offnuc}, top panel. The AGN are observed to have a lower SFR distribution while the SF-AGN, Composite and LINER sources present higher SFR values. Similar distributions were also found for the offsets in Fig.~\ref{offset}. This suggests that the AGN emission in the central region is more diluted by star formation in the SF-AGN, Composite and LINER sub-samples. The median of the central SFR for the AGN (log SFR = -3.6 $\pm$ 0.7 M$_{\odot}$ yr$^{-1}$) is slightly larger than for the SF-AGN (-3.3 $\pm$ 0.6 M$_{\odot}$ yr$^{-1}$), Composite (-3.1 $\pm$ 0.6 M$_{\odot}$ yr$^{-1}$) and LINERs (-3.4 $\pm$ 0.7 M$_{\odot}$ yr$^{-1}$), and applying a two-sample Kolmogorov-Smirnov test between the AGN and the SF-AGN/Composite/LINERs sub-samples yields p-values smaller than 0.05 in all cases, rejecting the null hypothesis that the data are drawn from the same distribution when using a confidence level of 95\%.

Similar results are found when checking if there are any differences in the central SFRs when dividing the sample into `nuclear' and `off-nuclear' AGN. As shown in Fig.~\ref{SFR3arcsec_nuc_offnuc}, bottom panel, the nuclear sources tend to have lower values of SFRs, indicative of a lower AGN dilution. The median value for the nuclear AGN (log SFR = -3.9 $\pm$ 0. M$_{\odot}$ yr$^{-1}$) is indeed lower than for the off-nuclear AGN (log SFR = -3.3 $\pm$ 0.6 M$_{\odot}$ yr$^{-1}$), and a two-sample Kolmogorov-Smirnov test rejects the null hypothesis that the data are drawn from the same distribution with a p-value $\ll$ 0.05. This indicates that dilution by star formation can be one of the causes of the offsets observed in Fig.~\ref{offset}, though many of the sources could still be off-nuclear AGN or switched-off AGN. 

\begin{figure}
\centering
\includegraphics[width=0.49\textwidth]{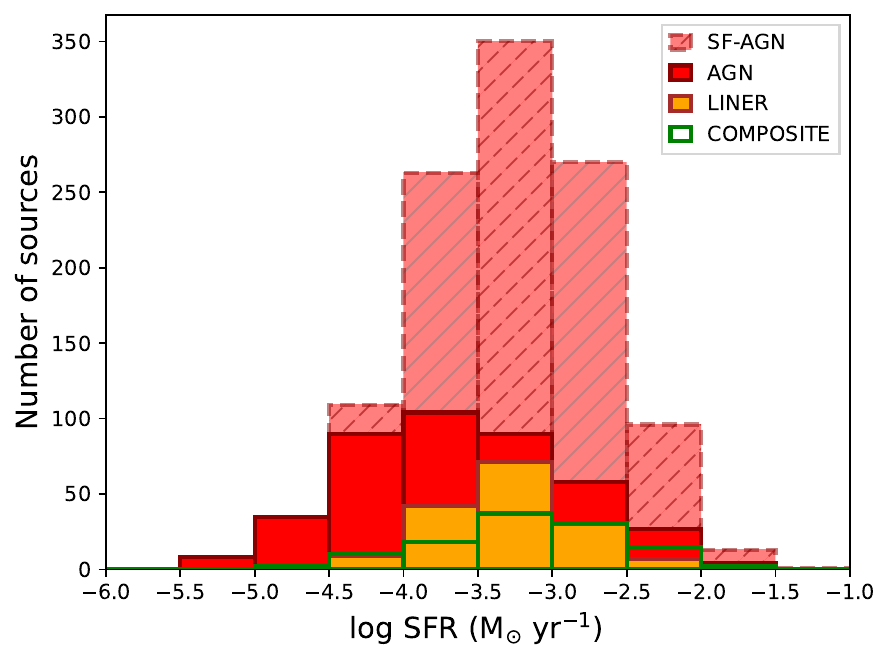}
\includegraphics[width=0.49\textwidth]{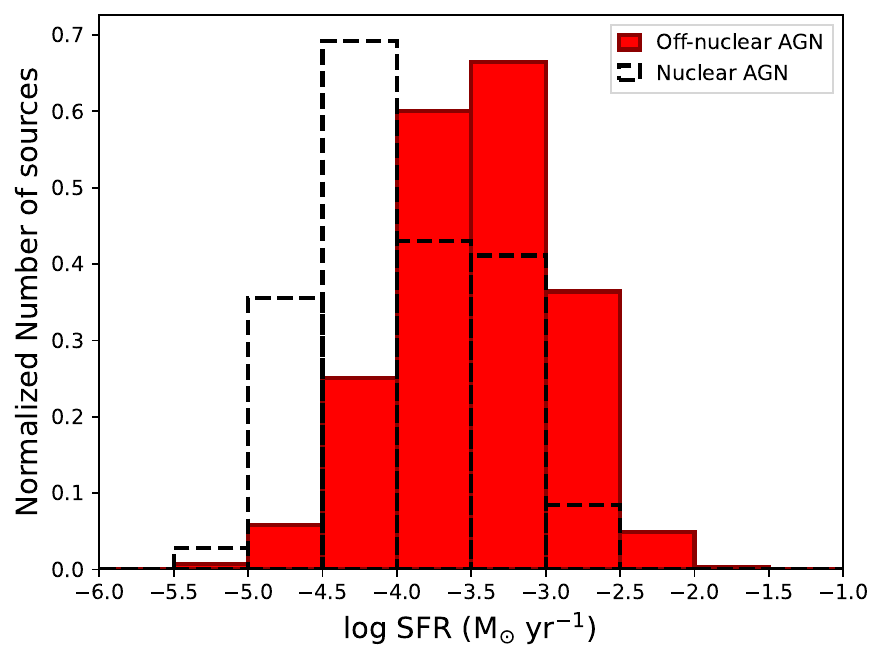}
\caption{Top: Distribution of the SFR (corrected for extinction) within the central 3 arcsec for the different sub-samples. Bottom: Distribution of the SFR within the central 3 arcsec for the nuclear AGN (i.e. those with offsets $\leq$ 3 arcsec) and off-nuclear AGN (i.e. with offsets $>$ 3 arcsec).}
\label{SFR3arcsec_nuc_offnuc}
\end{figure}

To look for off-nuclear AGN we make use of the radio, X-ray, and mid-infrared counterparts found in Sect.~\ref{multiwavelength}. We compute the offset between the optical center of the galaxy and the X-ray, radio, and infrared position for those AGN with an X-ray, radio, and mid-infrared counterpart, respectively. We show in Fig.~\ref{Xrayoffset}, top panel, the optical versus X-ray offset for each of the four sub-samples. As it can be seen most of the X-ray offsets are of less than 3 arcsec, indicating that for these sources the AGN is indeed at the center and thus that the optical offset observed must be caused by AGN dilution in the optical regime (i.e. due to star formation). However, three of the X-ray AGN have both optical and X-ray offsets beyond the 3 arcsec fiber of the SDSS, suggesting these are off-nuclear AGN. Another off-nuclear source is also found in the mid-infrared (Fig.~\ref{Xrayoffset}, bottom panel). All the radio AGN, on the contrary, have radio offsets within 3 arcsec, indicating that the optical offsets are an artifact of AGN dilution (Fig.~\ref{Xrayoffset}, middle panel).
The finding of off-nuclear AGN in MaNGA was already suggested by MDS+2020 and has been found in other works both at X-ray (e.g., \citealt{2018MNRAS.478.2576M}) and radio wavelengths (e.g., \citealt{2020ApJ...888...36R}), in agreement with simulations predicting that a vast population of IMBHs should be found as off-nuclear wanderers in dwarf galaxies (e.g., \citealt{2019MNRAS.482.2913B}; \citealt{2019MNRAS.486..101P}; \citealt{2023MNRAS.525.1479D}).

Nuclear stellar clusters (NSCs) have also been often found to be offset from the dwarf galaxy photocenter (e.g.,  \citealt{2006ApJS..165...57C}; \citealt{2021MNRAS.506.5494P}). In dwarf galaxies with M$_\mathrm{*} \sim 10^{9}$ M$_{\odot}$ the nucleation fraction is of $\sim$90\% (\citealt{2019ApJ...878...18S}) and the NSCs do not replace but coexist with the central BHs (e.g., \citealt{2006ApJ...644L..21F}; \citealt{2008ApJ...678..116S}). This could be also the case for the AGN dwarf galaxies here reported, for which the offset between the AGN and the optical center of the galaxy is in the same range as that of the offset between the NSC and the dwarf photocenter of \citeauthor{2021MNRAS.506.5494P} (2021, NSC offset= 0.01-7.26 arcsec with a median value of 0.74 arcsec).

\begin{figure}
\centering
\includegraphics[width=0.49\textwidth]{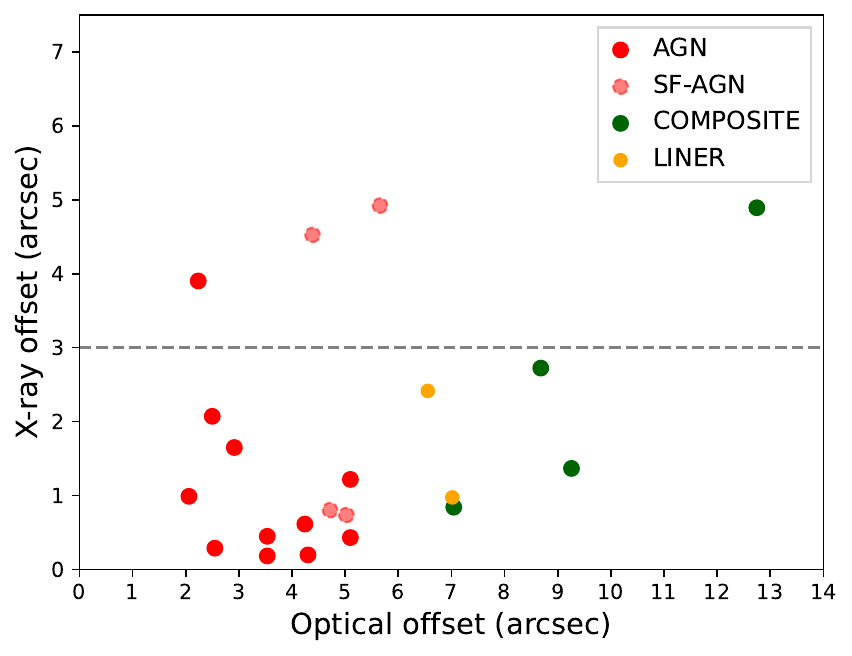}
\includegraphics[width=0.49\textwidth]{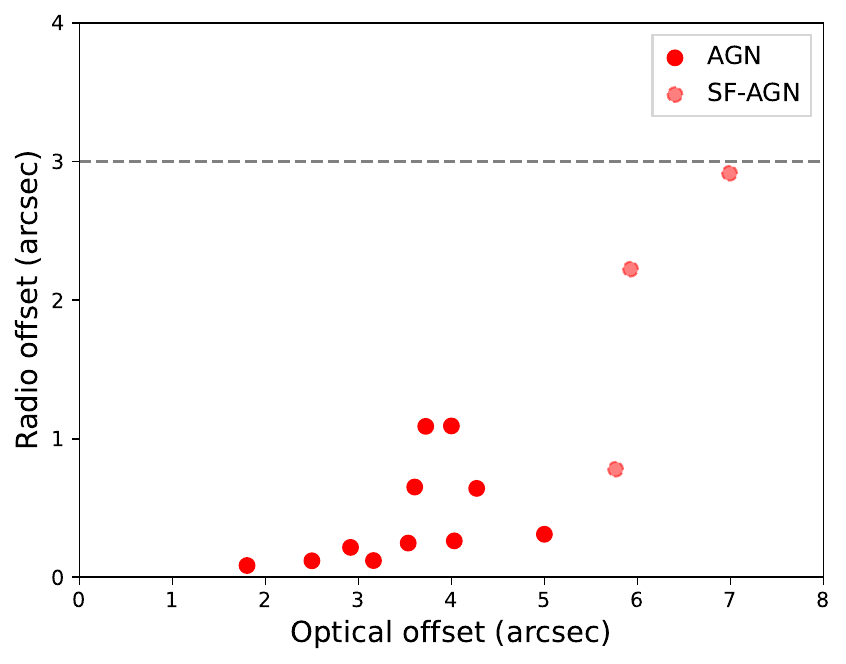}
\includegraphics[width=0.49\textwidth]{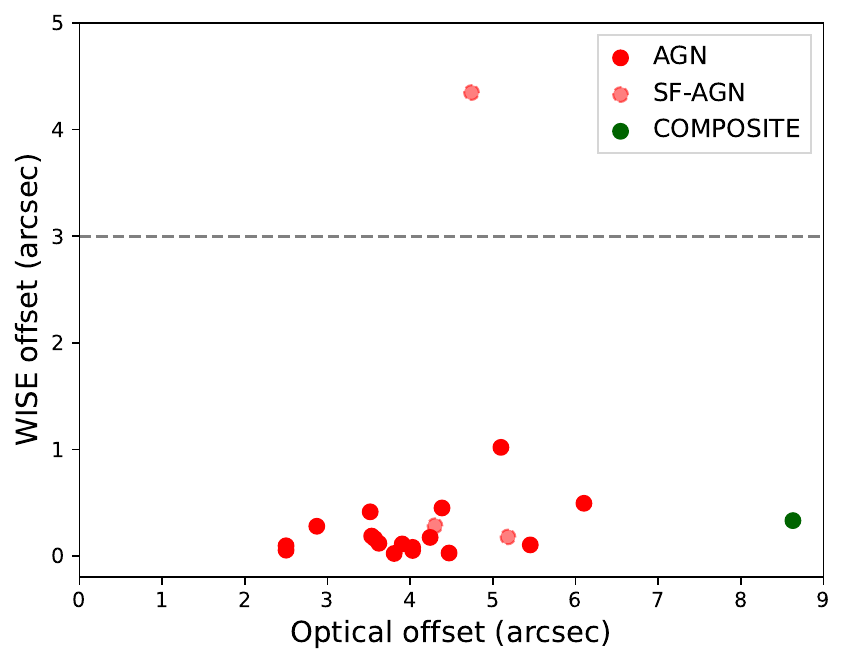}
\caption{Optical offset versus offset between the optical center of the galaxy and the X-ray position (top panel), the radio position (middle panel), and the WISE position (bottom panel) for those AGN, SF-AGN, LINER and Composite sources with an X-ray, radio, and WISE counterpart, respectively. The dashed horizontal line marks the SDSS central spectroscopic fiber of 3 arcsec.}
\label{Xrayoffset}
\end{figure}

\subsection{Galaxy properties}
We retrieve the photometric and morphological properties of the dwarf galaxies from the MaNGA PyMorph photometric Value Added Catalogue (MPP-VAC-DR17) and the MaNGA Deep Learning Morphological Value Added Catalogue (MDLM-VAC-DR17) of \cite{2022MNRAS.509.4024D}. The MPP-VAC-DR17 provides S\'ersic and S\'ersic+Exponential fits to the 2D surface brightness profiles of the final MaNGA DR17 galaxy sample in the g, r, and i bands, while the MDLM-VAC-DR17 provides the morphological classifications based on Deep Learning for the same set of galaxies. 
Out of the 2,362 dwarf galaxies, 2,357 have photometric fits and morphological classifications available in the MPP-VAC-DR17 and MDLM-VAC-DR17 catalogues, respectively. Of these, 2,048 have T-Type $>$0 (consistent with a late-type classification) and 309 T-Type $<$0 (consistent with an early-type classification) in the MPP-VAC-DR17, thus $\sim$87\% of the dwarves can be classified as late-type galaxies (LTGs). This fraction drops to 70\% when using the complementary classification based on P$_{\rm LTG}$, the probability of being late-type rather than early-type, and setting P$_{\rm LTG}$ > 0.5 as a threshold for selecting LTGs. This difference probably corresponds to faint and small galaxies for which the spiral features may not be so evident (see Figure 6 in \citealt{2022MNRAS.509.4024D}).

A fit to the surface brightness profile is available for 2,280 of the dwarves, of which 1,844 ($\sim$81\%) are best fitted with a S\'ersic profile of index $\leq$2 or with a S\'ersic+Exponential profile (see Fig.~\ref{histphotometry}, top panel), consistent with the morphological classification provided by the MDLM-VAC, according to which most of the galaxies in the sample are late-type.

From the g- and r-band photometry provided in the MPP-VAC-DR17 catalogue we also derive the B-V color based on the \cite{2005AJ....130..873J} transformations. We find a median value of <B-V> = 0.6 $\pm$ 0.2 (see Fig.~\ref{histphotometry}, bottom panel), fully consistent with that already found by MDS+2020 for their smaller sample of 37 AGN dwarf galaxies and redder than that of AGN dwarf galaxies identified from SDSS spectroscopy (e.g., \citealt{2013ApJ...775..116R}) or from X-ray and radio emission (e.g., \citealt{2018MNRAS.478.2576M,2019MNRAS.488..685M}; \citealt{2020ApJ...888...36R}). The galaxy colors are still <B-V> = 0.6-0.7 when distinguishing between the AGN, the SF-AGN, Composite and LINERs, indicating no differences in galaxy type among these sub-samples.

\begin{figure}
\centering
\includegraphics[width=0.49\textwidth]{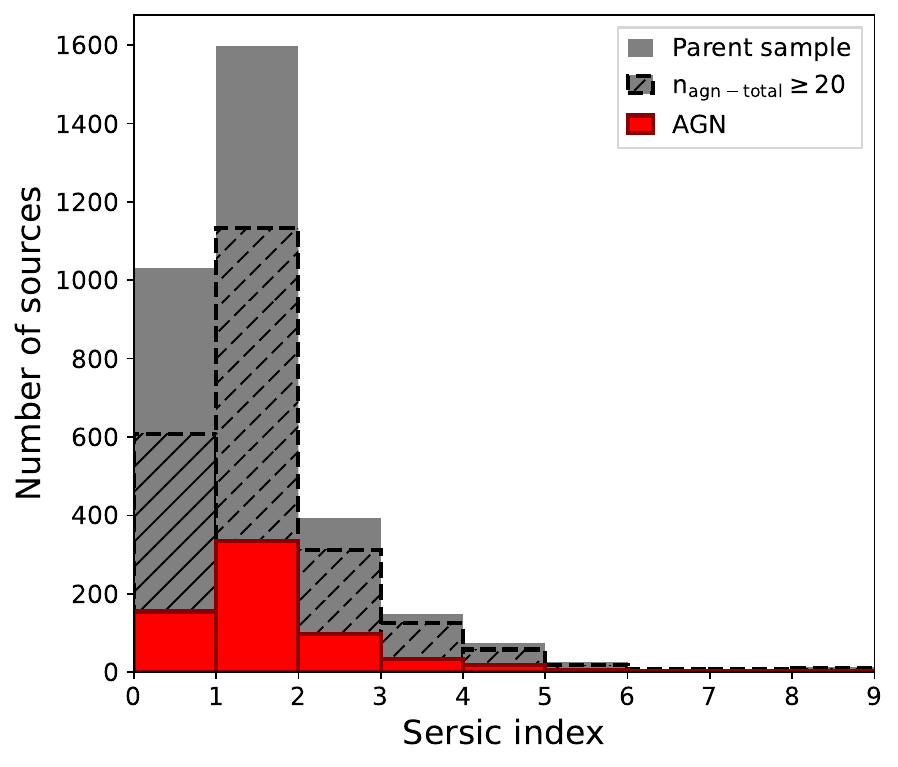}
\includegraphics[width=0.49\textwidth]{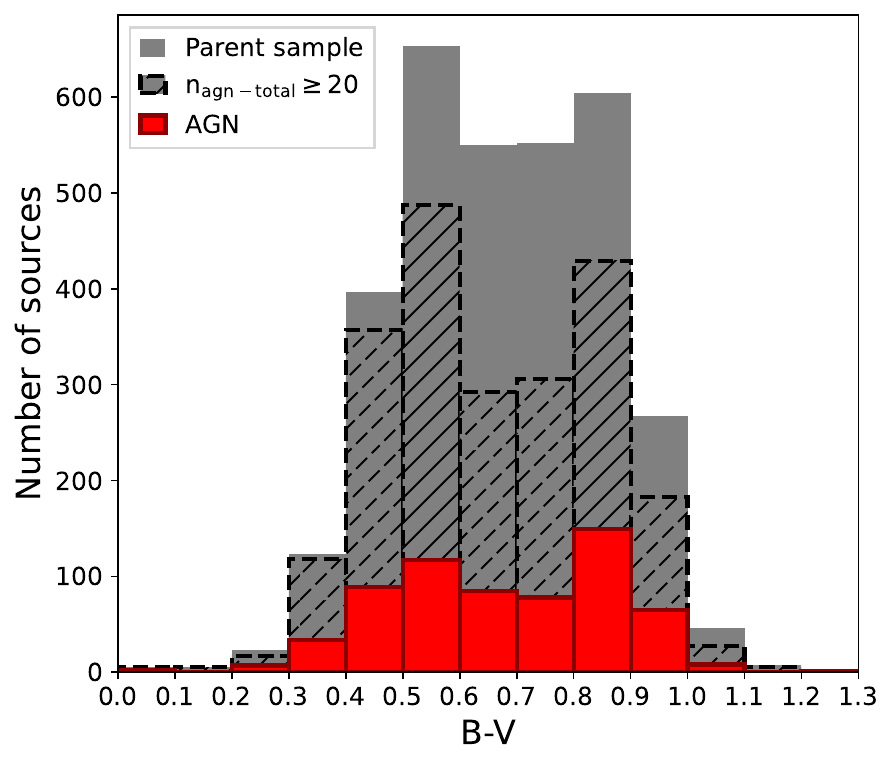}
\caption{Distribution of the S\'ersic index (top) and B-V color (bottom)  for the parent sample of 3,400 unique galaxies, the sample of 2,362 galaxies with strong indication of AGN ionisation, and the sample of 664 AGN candidates.} 
\label{histphotometry}
\end{figure}

\subsection{AGN demographics}
Albeit completeness, we derive the fraction of dwarf galaxies that host an AGN based on the classification scheme applied in Sect.~\ref{AGNclassification}. For this we use the initial sample of dwarf galaxies (3,436) but after removing the 36 duplicates, this is, 3,400 dwarf galaxies. 
Considering the sample of 664 AGN dwarf galaxies we obtain an AGN fraction of $\sim$20\%. %(664/3400*100).
This is significantly larger than any previous studies using integrated spectra (AGN fraction $<$1\%; e.g., \citealt{2013ApJ...775..116R}; \citealt{2022ApJ...937....7S}), MaNGA IFU ($\lesssim$5\%; e.g., \citealt{2018MNRAS.474.1499W}; MDS+2020) or corrected for completeness in the X-rays ($<$1\%; e.g., \citealt{2018MNRAS.478.2576M}; \citealt{2020MNRAS.492.2268B,2022MNRAS.510.4556B}). The AGN fraction would double to $\sim$54\% when considering both the AGN and the 1,176 SF-AGN, as found also by \cite{2022ApJ...931...44P} ($\sim$16-30\% when they include their SF-AGN), disfavoring direct collapse as the predominant seed BH formation model. We note that adding the 122 Composite sources the AGN fraction would increase even more, to $\sim$58 \%. We do not take the LINER sources into account given the possibility that their emission is caused by shocks, although we find that the median of the [OI]$\lambda$6300/H$\alpha$ ratios are in general smaller than 0.1 (above which shocks are the main excitation mechanism of the [OI]$\lambda$6300 line, e.g., \citealt{2021MNRAS.501L..54R}) and the [SII]$\lambda$6717,6731/H$\alpha$ ratio is smaller than 0.4 (above which the ionisation can be caused by supernova remnant shocks, e.g., \citealt{1980A&AS...40...67D}).

We note that the AGN identification approach applied in this paper recovers most of the AGN dwarf galaxies found in previous works using MaNGA: The six AGN dwarves from \cite{2018MNRAS.476..979P} are classified in our sample as AGN. Of the 48 AGN dwarves with $M_\mathrm{*} < 3 \times 10^{9}$ M$_{\odot}$ found by \citeauthor{2018MNRAS.474.1499W} (2018; see MDS+2020) we recover here 36 sources, classified according to our scheme as AGN (22), Composite (5), and SF-AGN (4). Five of the remaining AGN from \cite{2018MNRAS.474.1499W} are classified by us as Retired when applying the WHAN diagram, while the remaining ones do not pass the threshold of $n_\mathrm{agn-total} \geq$ 20 (see Fig. \ref{flowchart}).
Of the 37 AGN dwarf galaxies from MDS+2020 we recover 35 sources: 28 AGN, 5 Composite, and 1 SF-AGN, while the remaining one is here classified as Retired by the WHAN diagram. The remaining two sources do not have $n_\mathrm{agn-total} \geq$ 20.
\cite{2020ApJ...901..159C} perform an AGN identification using multiwavelength (X-ray, radio, and infrared) diagnostics, rather than optical emission line ratios, of a sample of 6,261 MaNGA galaxies. They report 406 AGN, of which 28 are classified as dwarf galaxies and are included in our sample. Of these 28 sources in common, 11 qualify as AGN, six as SF-AGN, nine as Composite, and one as LINER according to our diagnostics. We thus recover most of the \cite{2020ApJ...901..159C} AGN dwarf galaxies when using our spaxel by spaxel emission line ratio classification scheme. 

\subsection{AGN luminosity, black hole mass, and accretion rate}
As we have seen in Sect.~\ref{SDSS}, the use of MaNGA IFU allows us to recover $\sim$80\% more AGN than using integrated SDSS spectra, which can be partially explained by star-formation dilution. In this section we investigate any effects caused by the AGN bolometric luminosity and accretion rate. 
We derive the bolometric luminosity for the sub-sample of AGN from the median of the [OIII] luminosity of those spaxels classified as either AGN or Composite in the [NII]-BPT (e.g., MDS+2020) and assuming a bolometric correction factor of 1000 (\citealt{2014AJ....148..136M}). In Fig.~\ref{histLbol} we show the distribution of bolometric luminosities ($L_\mathrm{bol}$) for all the AGN (in red), only for the AGN classified also as AGN when using the SDSS spectra (in purple), and for a sample of X-ray selected AGN in dwarf galaxies from \citeauthor{2018MNRAS.478.2576M} (2018, in blue). The AGN are found to have in all cases log $L_\mathrm{bol} < 10^{42}$ erg s$^{-1}$, consistent with low-luminosity AGN (e.g., \citealt{2014ApJ...787...62M}) and lower than those of X-ray selected (and also radio selected, see MDS+2020) AGN dwarves.

The distribution of $L_\mathrm{bol}$ for the whole sub-sample of AGN reaches, in addition, lower values than those AGN that are SDSS AGN, as already found by MDS+2020. Indeed, the $L_\mathrm{bol}$ ranges from log $L_\mathrm{bol}$ = 38.9 to 41.6 erg s$^{-1}$ for those AGN that are also SDSS AGN, while log $L_\mathrm{bol}$ = 38.2 - 40.6 erg s$^{-1}$ for those AGN missed by the SDSS. Those sources unidentified as AGN by the integrated SDSS spectra seem thus to be fainter AGN. 
To further investigate whether this is related with a lower BH accretion rate we derive the Eddington ratio ($\lambda_\mathrm{Edd}$) as $\lambda_\mathrm{Edd} = L_\mathrm{bol}/(M_\mathrm{BH} \times 1.3 \times 10^{38})$. Since most of the AGN do not show broad emission lines (see next Section), we derive $M_\mathrm{BH}$ using the $M_\mathrm{BH}$-$M_\mathrm{*}$ correlation of AGN derived by \cite{2015ApJ...813...82R}. This yields a range of $M_\mathrm{BH}$ for the AGN of log $M_\mathrm{BH}$ = 4.5 - 5.9 $M_\mathrm{\odot}$ (with a typical uncertainty of 0.5 dex), consistent with being IMBHs. Based on these BH masses, the Eddington ratio for the AGN that are also AGN in SDSS is found to range from $\lambda_\mathrm{Edd} \sim 2 \times 10^{-5}$ to 2 $\times 10^{-2}$ while $\lambda_\mathrm{Edd} \sim 7 \times 10^{-6} - 2 \times 10^{-3}$ for those AGN missed by the SDSS. The use of MaNGA therefore allow us to identify a population of lower Eddington ratio sources than with SDSS. 
We note that deriving the BH mass from the $M_\mathrm{BH}$-$\sigma_{*}$ correlation of \cite{2018ApJ...855L..20M} or \cite{2020ARA&A..58..257G}, where $\sigma_{*}$ is the flux-weighted mean stellar velocity dispersion of all spaxels within 1 $R_\mathrm{eff}$ provided by the DAP, yields Eddington ratios that are in all cases below $\sim10^{-2}$. Hence all the AGN dwarf galaxies identified by MaNGA are accreting at sub-Eddington rates.

\begin{figure}
\centering
\includegraphics[width=0.49\textwidth]{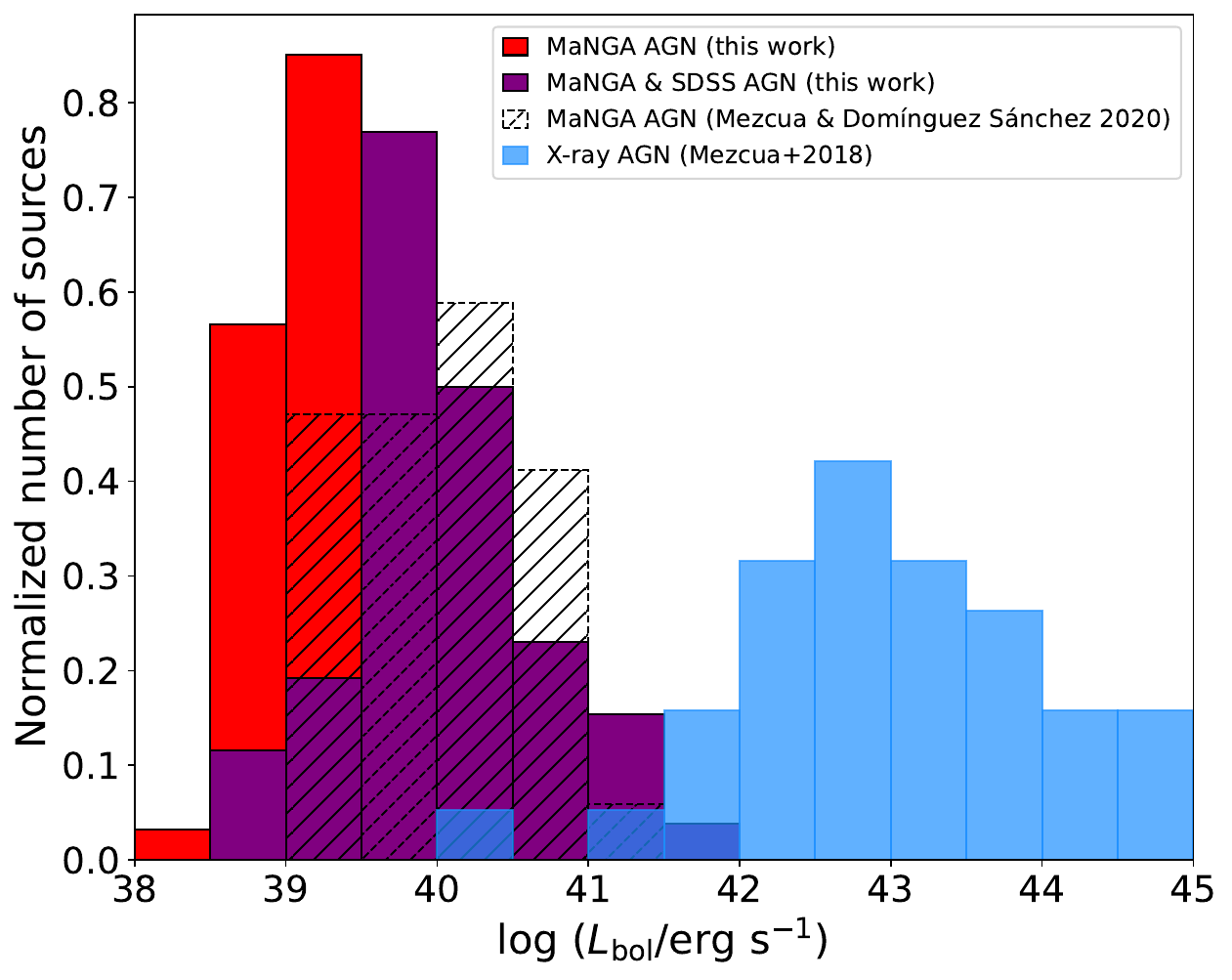}
\caption{Distribution of the bolometric luminosity for the AGN (red bars), the AGN that are also identified as AGN by SDSS (purple bars), the sample of 37 AGN dwarf galaxies from MDS+2020 (hashed bars), and the X-ray sample of AGN dwarf galaxies from \citeauthor{2018MNRAS.478.2576M} (2018, blue bars).}
\label{histLbol}
\end{figure}

\begin{table*}
\footnotesize{}
\caption{Broad-line AGN among the sample of 2,362 dwarf galaxies}
\label{type1AGN}
\begin{tabular}{{lccccccccl}}
\hline
\hline 
MaNGA & RA & DEC & $z$ &  log $M_\mathrm{*}$ & Type & $FWHM_\mathrm{H\alpha broad}$ & log $L_\mathrm{H\alpha broad}$ & log $M_\mathrm{BH}$ & Ref.\\
plateifu & (J2000) & (J2000) &  & (M$_{\odot}$) &  & (km s$^{-1}$) & (erg s$^{-1}$) & (M$_{\odot}$) & \\
(1) & (2)   & (3)   & (4)   & (5)   & (6)   & (7)   &  (8)  & (9)   & (10)\\
\hline
10223-3702 &  33.524643 & -0.276961 &  0.037 &          9.2 &    AGN &    2223 $\pm$ 241 &       38.2 $\pm$ 0.1 &        5.5 &          Thiswork \\
10493-1902 & 124.327367 & 52.029915 &  0.039 &          9.2 &    AGN &     1408 $\pm$ 55 &        40.9     &    6.4 &          \cite{2019ApJS..243...21L} \\
11980-1901 & 255.494175 & 24.236424 &  0.041 &          9.3 &    AGN &       -- &          -- &        5.0 & \cite{2018ApJ...863....1C} \\
12483-9102 & 186.453995 & 33.546902 &  0.001 &          8.8 &    AGN &       -- &          -- &        5.0 & \cite{2018ApJ...863....1C} \\
12622-1902 & 202.331963 & 32.004722 &  0.024 &          9.3 &    AGN &    2538 $\pm$ 360 &       37.8 $\pm$ 0.1 &        5.4 &          Thiswork \\
 8252-1902 & 146.091838 & 47.459851 &  0.026 &          8.8 &    AGN &       -- &          -- &        5.1 & \cite{2018ApJ...863....1C} \\
 8442-1901 & 199.675905 & 32.918662 &  0.036 &          9.2 &    AGN &    1131 $\pm$ 377 &       38.6 $\pm$ 0.2 &        5.1 &          MDS+2020 \\
 8446-1901 & 205.753337 & 36.165656 &  0.024 &          9.2 &    AGN &     1321 $\pm$ 99 &        40.6   &      6.1 &          \cite{2019ApJS..243...21L} \\
 8565-1902 & 241.033211 & 47.349658 &  0.042 &          9.4 &    AGN &       -- &          -- &        8.6 &           \cite{2008ApJ...688..826L} \\
8626-12704 & 263.755219 & 57.052433 &  0.047 &          8.4 & SF-AGN &     771 $\pm$ 109 &       39.1 $\pm$ 0.1 &        5.0 &          Thiswork \\
 8982-3703$^{\dagger}$ & 203.190094 & 26.580376 &  0.047 &          8.9 &    AGN &     1854$\pm$52 &        39.12 $\pm$ 0.02 &        5.8 &          Thiswork \\
 8992-3702 & 171.657262 & 51.573041 &  0.026 &          9.5 &    AGN &     1674 $\pm$ 48 &        40.0    &     6.1 &         \cite{2019ApJS..243...21L} \\
 9000-1901 & 171.400654 & 54.382574 &  0.021 &          9.3 &    AGN &   3352 $\pm$ 1101 &        41.3   &    7.3 &          \cite{2019ApJS..243...21L} \\
 9048-6101 & 244.036041 & 24.079529 &  0.032 &          9.3 &    AGN &    3252 $\pm$ 681 &       37.9 $\pm$ 0.1 &        5.7 &          Thiswork \\
 9091-3704 & 241.944669 & 25.537511 &  0.041 &          9.4 &    AGN &    1461 $\pm$ 111 &        38.62 $\pm$ 0.04 &        5.3 &          Thiswork \\
 9190-1901 &  53.741686 & -5.814780 &  0.018 &          9.1 &    AGN &    4648 $\pm$ 135 &        39.9   &    7.0 &          \cite{2019ApJS..243...21L} \\
 9891-1901 & 227.177572 & 28.171156 &  0.026 &          9.4 &    AGN &    3778 $\pm$ 570 &       38.0 $\pm$ 0.1 &        5.9 &          Thiswork \\
\hline
\hline
\end{tabular}
\smallskip\newline\small {\bf Column designation:}~(1) MaNGA plateifu; (2,3) RA, DEC coordinates of the optical center of the galaxy or IFU center; (4) galaxy redshift; (5) galaxy stellar mass; (6) MaNGA BPT classification; (7, 8) FWHM and luminosity of the broad H$\alpha$ emission line component; (9) BH mass derived from the broad H$\alpha$ component; (10) Reference for columns 7, 8, 9. The typical uncertainties of the BH masses are of 0.5 dex. $^{\dagger}$ This source has also been recently reported as a broad-line AGN by \cite{2023MNRAS.524.5827F}. %$^{*}$Galaxy with a FIRST radio counterpart ( 9091-3704). 
\end{table*}

\subsection{Type 1 AGN}
\label{BHmass}
Eight of the AGN in the sample of 2,362 dwarf galaxies are found to have broad Balmer emission lines by \cite{2019ApJS..243...21L} and \cite{2018ApJ...863....1C}, see Table~\ref{type1AGN}. Another source, 8442-1901, was found to be a hidden type 1 AGN by MDS+2020, showing broad H$\alpha$ emission in the stacked spectrum of the AGN spaxels. 
We check whether any other hidden broad-line AGN are present among the sample of 2,362 dwarf galaxies by stacking the spectra of those spaxels classified as either AGN, AGN-WHAN or SF-AGN and fitting the resulting stacked spectrum with a multi-Gaussian Python routine. The details of the procedure can be found in  MDS+2020 (Appendix A). In short,  for each galaxy, we stack together all the spaxels belonging to the class AGN, AGN-WHAN or SF-AGN using the LOGCUBE-HYB10-MILESHC-MASTARHC2 from MaNGA DR17.\footnote{This is a small variation with respect to MDS+2020, where we used the HYB10-GAU-MILESHC cubes.}The stacked spectrum is the median value of the 3$\sigma$ clipped and normalized rest-framed flux at each wavelength, after removing masked regions or skylines.} 
The H$\alpha$ + [NII] emission line complex of the stacked AGN spectrum is fitted assuming the same width for the [NII]$\lambda$6583, [NII]$\lambda$6548, and narrow H$\alpha$ emission line Gaussian components and allowing for a secondary broad H$\alpha$ component. The flux ratio of the [NII] is set to the theoretical value of 3.06 and the relative separation between the centers of the narrow components is fixed to their theoretical wavelengths. Those sources best-modeled (i.e. lower $\chi$$^{2}$) with a broad and narrow H$\alpha$ emission line component are subject to visual inspection, based on which we confirm the good fit of the complex or remove ambiguous cases. This yields 8 broad-line sources. One of them (8565-1902) is reported to have broad H$\beta$ emission by \cite{2008ApJ...688..826L}.
At the time of writing, \cite{2023MNRAS.524.5827F} reported a catalogue of type 1 AGN in the MaNGA survey. Four of their sources are among our sample of dwarf galaxies, three of which are already included in the \cite{2019ApJS..243...21L} sample. The forth one we report it here as a new broad-line AGN (8982-3703).
%This yields 12 broad-line sources, 4 of which are already included in the samples of Liu et al. (2019) and Chilingarian et al. (2018). We also recover one additional type 1 AGN (8565-1902) reported to have broad H$\beta$ emission by Li, Wu, and Wang (2008). 
The remaining 6 sources do not have (to the best of our knowledge) any broad Balmer emission reported in the literature and are thus reported here as new type 1 AGN in dwarf galaxies. The main properties of the in total new 7 sources (when including 8982-3703) are shown in Table~\ref{type1AGN}. They are all AGN according to our emission line classification criteria except for 8626-12704, which is a SF-AGN. The FWHM of their broad H$\alpha$ emission line component, corrected for instrumental resolution, thermal and natural broadening, ranges from $\sim$772 to 3,779 km s$^{-1}$. Based on these FWHM and the luminosity of the broad H$\alpha$ emission line component we derive a range of BH masses of log $M_\mathrm{BH}$ = 5.0 - 5.9 $M_\mathrm{\odot}$ using eq. 5 in \cite{2013ApJ...775..116R}, which favors an IMBH nature.

\section{Conclusions}
\label{conclusions}
In this paper we have made use of the final MaNGA data release (DR17) to search for AGN in dwarf galaxies using a spaxel by spaxel classification scheme based on three spatially resolved emission line diagnostic diagrams ([NII]-, [SII]-, and [OI]-BPT) plus the WHAN diagram. This has resulted in a sample of 664 AGN, 122 Composite, and 173 LINERs among a parent sample of 3,400 dwarf galaxies. We additionally identify 1,176 SF-AGN, a classification recently introduced by \cite{2022ApJ...931...44P} for those sources which are SF in the [NII]-BPT but AGN in the [SII]- and [OI]-BPT. Nearly all of these SF-AGN and 80\% of the AGN are not identified as AGN when using single-fiber SDSS emission line diagnostics. This can be explained by the finding that the offset between the optical center of the galaxy and the median position of the AGN spaxels is larger than the 3-arcsec central fiber of the SDSS for $\sim$98\% of the SF-AGN and $\sim$62\% of the AGN. The offset of the SF-AGN, Composite and LINER sub-samples are in general found to be slightly larger than that of the AGN. Such differences can be in part attributed to dilution by star formation, though many of the sources could also be either switched-off AGN or off-nuclear AGN, specially those with optical offsets above 3 arcsec. Based on the offset between the X-ray, radio, or mid-infrared counterpart and the optical center of the galaxy, we are able to confirm the off-nuclear AGN nature for four sources (three SF-AGN and one Composite; see Sect.~\ref{multiwavelength}). This small fraction is largely caused by the reduced sky coverage and sensitivity of the multiwavelength surveys here used. A high fraction ($\sim$50\%) of off-nuclear AGN in dwarf galaxies is expected from simulations (e.g., \citealt{2019MNRAS.482.2913B}; \citealt{2019MNRAS.486..101P}), and several cases have been indeed already found observationally before (e.g., \citealt{2018MNRAS.478.2576M}; \citealt{2020ApJ...888...36R}). Further observations (e.g. in the X-ray regime) should increase the number of confirmed off-nuclear AGN. 

When considering only the sub-sample of 664 AGN, we find that those that are not identified as AGN by SDSS spectroscopy have lower accretion rates than those that are AGN both in SDSS and MaNGA, indicating that with MaNGA we are able to identify a population of lower Eddington ratio AGN in dwarf galaxies than with SDSS. By stacking those spaxels classified as AGN, we are, in addition, able to identify seven new broad-line AGN not identified as such in previous SDSS studies. For these seven sources we derive the BH mass from the width and luminosity of the broad H$\alpha$ component, while for the remaining broad-line AGN we use the $M_\mathrm{BH}$-$M_\mathrm{*}$ correlation of \cite{2015ApJ...813...82R}. We find a range of BH masses of log $M_\mathrm{BH}$ = 4.5 - 5.9 $M_\mathrm{\odot}$, consistent with IMBHs, and that all the sources are accreting a sub-Eddington rates.

Based only on the sub-sample of 664 AGN, the AGN fraction in dwarf galaxies is of $\sim$20\%, which is significantly larger than that found in single-fiber spectroscopic studies and X-ray surveys (i.e. $<$1\%; e.g., \citealt{2013ApJ...775..116R}; \citealt{2018MNRAS.478.2576M}; \citealt{2020MNRAS.492.2268B,2022MNRAS.510.4556B}; \citealt{2022ApJ...937....7S}). The use of IFU thus allows us to identify a large and so far hidden population of faint and low accretion rate AGN in dwarf galaxies, with BH masses in the IMBH regime. The addition of the SF-AGN would increase the AGN fraction to $\sim$54\%, which can have strong implications for understanding which was the predominant seed BH formation mechanism in the early Universe. 
  
\section*{Acknowledgments}
The authors thank the anonymous referee for insightful comments. 
M.M. acknowledges support from the Spanish Ministry of Science and Innovation through the project PID2021-124243NB-C22. This work was partially supported by the program Unidad de Excelencia Mar\'ia de Maeztu CEX2020-001058-M. HDS acknowledges the financial support from the Spanish Ministry of Science and Innovation and the European Union - NextGenerationEU through the Recovery and Resilience Facility project ICTS-MRR-2021-03-CEFCA and financial support provided by the Governments of Spain and Arag\'on through their general budgets and the Fondo de Inversiones de Teruel.

Funding for the Sloan Digital Sky 
Survey IV has been provided by the 
Alfred P. Sloan Foundation, the U.S. 
Department of Energy Office of 
Science, and the Participating 
Institutions. 

SDSS-IV acknowledges support and 
resources from the Center for High 
Performance Computing  at the 
University of Utah. The SDSS 
website is www.sdss4.org.

SDSS-IV is managed by the 
Astrophysical Research Consortium 
for the Participating Institutions 
of the SDSS Collaboration including 
the Brazilian Participation Group, 
the Carnegie Institution for Science, 
Carnegie Mellon University, Center for 
Astrophysics | Harvard \& 
Smithsonian, the Chilean Participation 
Group, the French Participation Group, 
Instituto de Astrof\'isica de 
Canarias, The Johns Hopkins 
University, Kavli Institute for the 
Physics and Mathematics of the 
Universe (IPMU) / University of 
Tokyo, the Korean Participation Group, 
Lawrence Berkeley National Laboratory, 
Leibniz Institut f\"ur Astrophysik 
Potsdam (AIP),  Max-Planck-Institut 
f\"ur Astronomie (MPIA Heidelberg), 
Max-Planck-Institut f\"ur 
Astrophysik (MPA Garching), 
Max-Planck-Institut f\"ur 
Extraterrestrische Physik (MPE), 
National Astronomical Observatories of 
China, New Mexico State University, 
New York University, University of 
Notre Dame, Observat\'ario 
Nacional / MCTI, The Ohio State 
University, Pennsylvania State 
University, Shanghai 
Astronomical Observatory, United 
Kingdom Participation Group, 
Universidad Nacional Aut\'onoma 
de M\'exico, University of Arizona, 
University of Colorado Boulder, 
University of Oxford, University of 
Portsmouth, University of Utah, 
University of Virginia, University 
of Washington, University of 
Wisconsin, Vanderbilt University, 
and Yale University.

\section*{Data availability}
The data underlying this article are available in the article and in its online supplementary material.


\bibliographystyle{mnras}
\bibliography{referencesALL}
%\bibliography{/Users/mmezcua/Documents/referencesALL}

%%%%%%%%%%%%%%%%%%%%%%%%%%%%%%%%%%%%%%%%%%%%%%%%%%
%%%%%%%%%%%%%%%%%%%%%%%%%%%%%%%%%%%%%%%%%%%%%%%%%%

\appendix

\section{Effects of SNR threshold selection}
\label{SNR}
\begin{figure}
\centering
\includegraphics[width=0.49\textwidth]{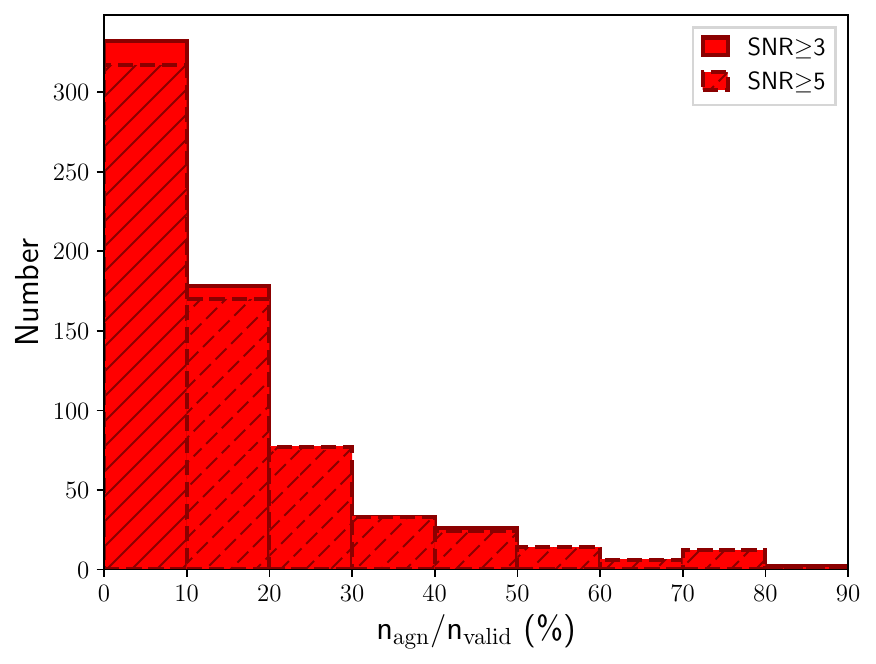}
\includegraphics[width=0.49\textwidth]{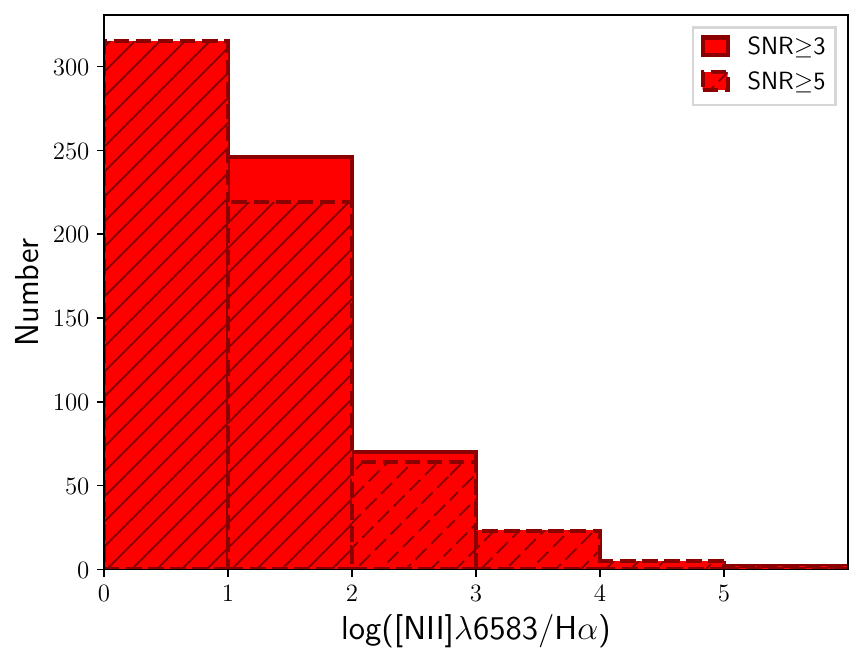}
\includegraphics[width=0.49\textwidth]{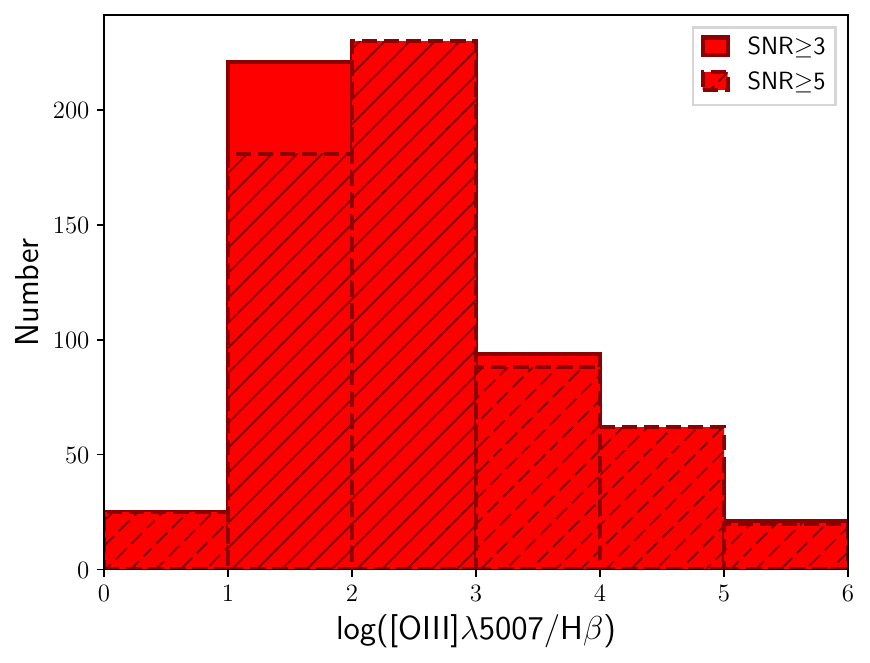}
\caption{Distribution of the percentage of $n_\mathrm{agn}/n_\mathrm{valid}$ spaxels (top), of the log([NII]$\lambda$6583/H$\alpha$) flux ratio (middle) and of the log([OIII]$\lambda$5007/H$\beta$) flux ratio (bottom) when using a threshold of SNR$\geq$3 and of SNR$\geq$5.}
\label{hist_SNRtests}
\end{figure}

\begin{figure*}
\centering
\includegraphics[width=0.49\textwidth]{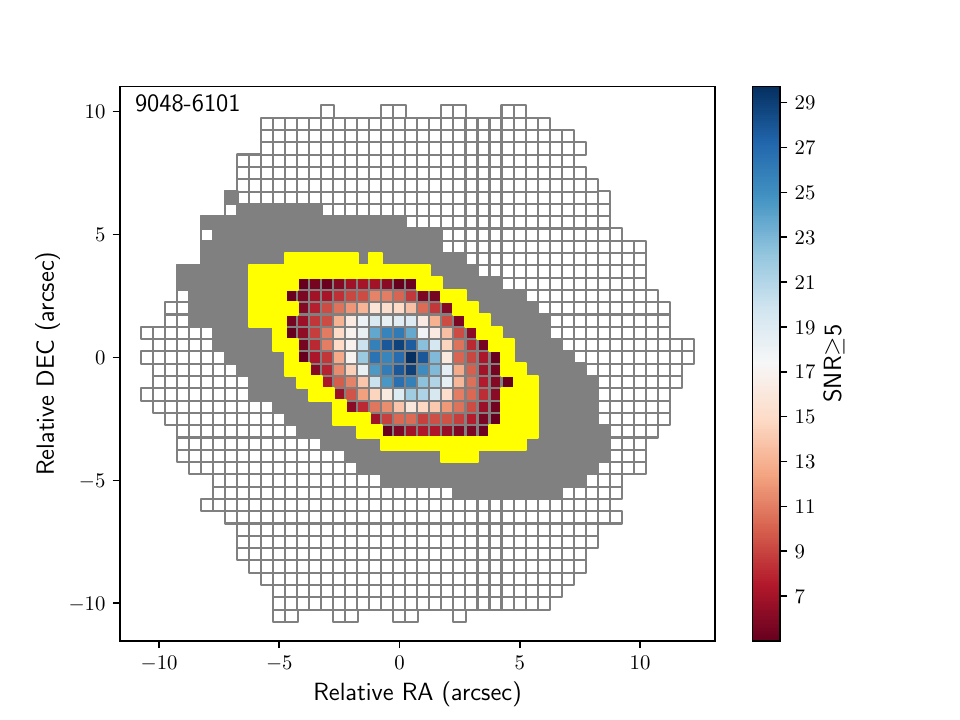}
\includegraphics[width=0.49\textwidth]{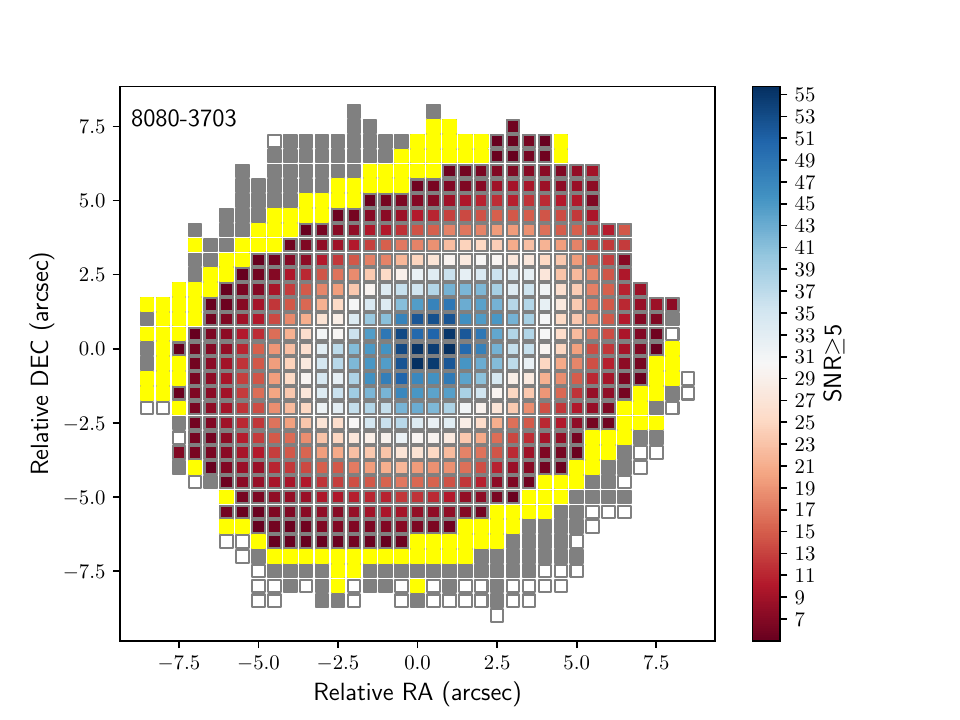}
\caption{Spatial distribution of the SNR for the same two examples as in Figs.~\ref{BPT-AGN}-\ref{BPT-SF-AGN}. The colorbar is set for those spaxels with SNR$\geq$5. Those spaxels with 3$\leq$SNR$<$5 are shown in yellow to highlight the number of spaxels that would be missed when applying a threshold of SNR$\geq$5 rather than SNR$\geq$3.}
\label{colorbarSNR}
\end{figure*}

To perform the spaxel classification in Sect.~\ref{spaxclass} we have considered only those spaxels whose BPT emission lines (H$\alpha$, H$\beta$, [NII]$\lambda$6583, [SII]$\lambda$6718+[SII]$\lambda$6732, [OIII]$\lambda$5007, [OI]$\lambda$6300) have SNR$\geq$3, as commonly adopted in BPT studies (e.g., \citealt{2013ApJ...775..116R} for SDSS; \citealt{2020MNRAS.492.4680W} for MaNGA; \citealt{2022ApJ...937....7S} for GAMA; \citealt{2023ApJ...954...77J} for SAMI). In Figs.~\ref{BPT-AGN}-\ref{BPT-SF-AGN} we have shown that the MaNGA DAP modeling is able to identify and fit the main BPT emission lines down to SNR=3. Here we test whether the number of AGN sources and BPT flux ratios would change when increasing the SNR threshold from SNR$\geq$3 to SNR$\geq$5. Based on the number of $n_\mathrm{agn}$/$n_\mathrm{valid}$ spaxels (defined in Sect.~\ref{AGNclassification}), when using SNR$\geq$5 rather than SNR$\geq$3 we still keep 99.7\% of the sources classified as AGN (see Fig.~\ref{hist_SNRtests}). 
The numbers are similar for the SF-AGN ($\sim$98\% of the sources). This can be also seen in the SNR distribution (see Fig.~\ref{colorbarSNR}) of the same two galaxies shown in Figs.~\ref{BPT-AGN}-\ref{BPT-SF-AGN}, where it can be observed that the main bulk of spaxels used for the galaxy classification have SNR$\geq$5. 
Based on the BPT flux ratios, when using SNR$\geq$5 rather than SNR$\geq$3 we still keep 96.6\% of the sources classified as AGN. We thus conclude that increasing the SNR threshold from SNR$\geq$3 to SNR$\geq$5 would not yield significant differences in the results reported in the paper.

% Don't change these lines
\bsp	% typesetting comment
\label{lastpage}
\end{document}